\documentclass[longauth]{aa}

\usepackage{graphicx}
\usepackage{lscape}
\usepackage{longtable}

\usepackage{txfonts}

\usepackage{longtable}
\usepackage{float}
\usepackage{subcaption}
\usepackage{siunitx}
\usepackage{multirow}
\usepackage{amsmath}
\usepackage{booktabs}
\usepackage{placeins}

\def\mstar  {${M_\star}$}
\def\teff  {$T_{\rm eff}$}
\def\macc   {${\dot{M}_{\rm acc}}$}
\def\lacc   {${L_{\rm acc}}$}
\def\mdust {${M_{\rm dust}}$}
\def\mdisc {${M_{\rm disc}}$}

\def\msun {${M_{\odot}}$}
\def\lsun {${L_{\odot}}$}
\def\lstar {${L_\star}$}

\newcommand{\Rco}{{$R_{\rm CO}$}}

\usepackage{color}

\usepackage[breaklinks,colorlinks,citecolor=blue]{hyperref}

\begin{document}

   \title{X-Shooter survey of disc accretion in Upper Scorpius\thanks{Based on observations collected at the European Southern Observatory under ESO programmes 097.C-0378(A), 0101.C-0866(A), 113.26NN.001, 113.26NN.003, 115.27XL.001, and 105.2082.003. }} 

\titlerunning{X-Shooter survey of disc accretion in Upper Scorpius}
\authorrunning{Empey et al.}

   \subtitle{II. A lack of correlation between accretion rates and disc properties}

   \author{A. Empey \inst{1}\thanks{Corresponding author: \texttt{aaron.empey@gmail.com}}, C.F. Manara \inst{2} \and R. Garcia Lopez \inst{1} \and A. Natta\inst{1,3} \and R. Claes\inst{4} \and F. Zagaria \inst{5,6} \and J.M. Alcalá \inst{7} \and R. Anania \inst{8} \and \\
   G. Beccari \inst{2} \and J. Carpenter \inst{9} \and S. Facchini \inst{10} \and D. Fedele \inst{11} \and G. Lodato \inst{10} \and K. Mauco \inst{2,12} \and A. Miotello \inst{2} \and B. Nisini \inst{13} \and \\
 I. Pascucci \inst{14} \and L. Piscarreta \inst{2} \and G. Rosotti \inst{10} \and A. Scholz \inst{15} \and L. Testi \inst{16} \and M. Vioque\inst{2}}

   \institute{University College Dublin (UCD), Belfiled, Dublin, Ireland  
   \and European Southern Observatory, Karl-Schwarzschild-Strasse 2, 85748 Garching bei M\"unchen, Germany  
   \and School of Cosmic Physics, Dublin Institute for Advanced Studies, 31 Fitzwilliams Place, Dublin 2, Ireland   
   \and Department of Physics, Maynooth University, Maynooth, Co. Kildare, Ireland 
   \and Max Planck Institute for Astronomy, Königstuhl 17, 69117 Heidelberg, Germany  
   \and Institue of Astronomy, University of Cambridge, Madingley Road, Cambridge CB3 OHA, UK  
   \and INAF-Osservatorio Astronomico di Capodimonte, via Moiariello 16, 80131 Napoli, Italy  
   \and School of Physics, Trinity College Dublin, the University of Dublin, College Green, Dublin 2, Ireland 
   \and Joint ALMA Observatory, Avenida Alonso de Córdova 3107, Vitacura, Santiago, Chile  
   \and Dipartimento di Fisica, Università degli Studi di Milano, Via Celoria 16, 20133 Milano, Italy  
   \and INAF-Osservatorio Astrofisico di Arcetri, Largo E. Fermi 5, I-50125 Firenze, Italy  
   \and Universidad Nacional Autónoma de México, Instituto de Astronomía, AP 106, Ensenada 22800, BC, México.  
   \and INAF—Osservatorio Astronomico di Roma, Via di Frascati, 33, 00078, Monte Porzio Catone, Italy  
   \and Lunar and Planetary Laboratory, the University of Arizona, Tucson, AZ 85721, USA  
   \and SUPA, School of Physics \& Astronomy, University of St Andrews, North Haugh, St Andrews KY16 9SS, UK 
   \and Alma Mater Studiorum – Università di Bologna, Dipartimento di Fisica e Astronomia “Augusto Righi”, Via Gobetti 93/2, I-40129,
Bologna, Italy \\  
}

 \date{Received 10 April 2026 / Accepted 16 June 2026}

  \abstract
  {The evolution of protoplanetary discs is intertwined with the process of planet formation, growth and migration. Studies of nearby star-forming regions of different ages and properties provide the information needed to understand the processes governing their evolution.}
  {This paper presents the results of a spectroscopic study of the stellar and accretion properties of a large sample of 127 stars with protoplanetary discs in the Upper Scorpius region, a relatively old (5-10 Myr), nearby ($\sim 145$pc) star-forming region, with disc dust masses inferred from ALMA continuum measurements.}
  {We derived the accretion luminosity from the excess UV continuum emission with respect to the photospheric and chromospheric emission self-consistently with the stellar spectral types, extinction, and luminosity, using the FitteR for Accretion ProPErties of T Tauri stars (FRAPPE) code. We applied a new method to evaluate upper limits on the accretion luminosity. In $\sim$50\% of cases, we could only evaluate upper limits on the accretion luminosity, either because the signal-to-noise ratio of the data was insufficient or because the measured value of the accretion luminosity was below the statistical estimate of the emission due to chromospheric activity.}
  {The mass accretion rate shows a weak correlation with stellar mass, while we find no correlation with disc properties such as dust mass or gaseous disc radius. The dispersion is larger than that found in younger star-forming regions such as Lupus and Chamaeleon I, and suggests fading of the correlations with age. We find no evidence that the observed dispersion can be explained by membership in Upper Scorpius sub-groups, or by the properties of known binary systems or transition discs.}
  {The lack of correlation and the large dispersion of accretion rates challenge the current expectations of evolutionary models. The observed properties point to a decoupling of the inner and outer discs by the age of Upper Scorpius and a fading of the relations observed in younger star-forming regions, which calls for further development of current theoretical frameworks to be explained.}


   \maketitle

\section{Introduction}
To date, thousands of exoplanets have been discovered as a result of advances in our observational capabilities. These discoveries have significantly improved our understanding of these worlds. However, fundamental unanswered questions remain regarding how they are born and evolve. A key component in tackling these issues is to better understand how their natal environments, the protoplanetary discs, evolve \citep{MorbidelliRaymond2016}. 

Disc properties evolve while planetary systems form and this evolution affects their final architecture and composition \citep{Drazkowska2023}. At the centre of this dramatic interplay is a young, evolving star that interacts with the disc through several accretion and ejection mechanisms critical to shaping the physical properties of the disc \citep{Hartmann2016, Pascucci2023}, until the disc itself dissipates \citep{Alexander2014}. By studying star-disc interactions, it is possible to study the mechanisms driving disc evolution. 

There remains a longstanding debate about the fundamental underlying mechanisms responsible for the evolution and eventual dispersal of these discs \citep{Manara2023}. On the one hand the viscous framework attributes disc evolution to the transport of angular momentum within the disc via turbulence-induced viscosity \citep{LindenbellPringle1974}. On the other hand, under the prescription of the magneto-hydrodynamic (MHD) wind-driven theory, angular momentum is instead removed via magneto-thermal winds \citep{BlandfordPayne1982,BaiStone2013,Bai2016,Lesur2021,Tabone2022}. In addition to these basic disc-evolution processes, many additional forces contribute to disc shaping and dispersal. These include external processes such as binary encounters or flybys \citep{ClarkePringle1993,Pfalzner2005,Cuello2023}, accretion of material from the environment \citep[e.g.][]{Gupta2023,Winter2024} and external photoevaporation \citep{Clarke2007, Anderson2013, Facchini2016, Winter&Haworth2022, Mauco2023}. 

The key to constraining such models and developing a more complete understanding of the disc evolutionary process lies in conducting demographic studies of different star-forming regions \citep{Lodato2017, Manara2023, Somigliana2024, Tabone2025}. Given the expected $\sim 3-10$ Myr lifetime of discs \citep[e.g.][]{Fedele2010, Ratzenbock_ages, Delfini2025}, studying regions of different characteristic ages acts as a tracer of the overarching evolutionary path. 

The key properties needed to constrain disc evolution mechanisms are the stellar mass (\mstar) and age, the mass accretion rate (\macc), as well as the bulk disc mass and size in gas and dust \citep[][]{Manara2023, Miotello2023}. These bulk properties are shown to evolve with time across various star-forming regions, reflecting their age \citep{Ansdell2017,Hendler2020}. Measurements of accretion rates and dust-based disc masses show a correlation in various star-forming regions \citep{Manara2016b,Mulders2017,Manara2020, Testi2022,AlmendrosAbad2024}. 
 
The physical mechanism to explain these dependences and how they evolve with time is still debated. 
Population-synthesis models show that the choice of evolutionary models (viscous, wind-driven, or hybrid, that is, a mixture of the two) has an impact on the dispersion in the relations and how they change with time \citep{Somigliana2024}. Notably, the wind-driven scenario appears more favourable for explaining the scatter in the observed correlations \citep{Somigliana2024}. Recently, JWST observations of MHD winds, which allow measurements of their mass-loss rates, offer a complementary approach for testing disc evolution models \citep{Pascucci2025}. Nonetheless, more observational surveys measuring accretion and disc properties are required to constrain the underlying mechanisms behind disc evolution and dispersal in these models. 

Measurements of the properties of young stars and their discs have become possible due to the advent of sensitive optical spectrographs such as X-Shooter \citep{vernet2011} on the European Southern Observatory's (ESO) Very Large Telescope (VLT) and millimetre interferometers - such as the Atacama Large Millimeter Array (ALMA), and the accurate distance measurements provided by the \textit{Gaia} satellite \citep{Gaia2016}. The availability of these powerful instruments paved the way for large observing campaigns to probe various nearby star-forming regions including Lupus \citep{Ansdell2016, Ansdell2018, Sanchis2020,Alcala2014,Alcala2017}, Corona Australis \citep{Cazzoletti2019}, Taurus \citep{Andrews2013, Akeson2019,Long2019}, Chamaeleon I \citep{Pascucci2016, Manara2016a, Manara2017}, and $\rho$ Ophiucus \citep{Cox2017,Cieza2019,Testi2022}. These have provided important catalogues of star and disc properties at young ages (1-3 Myr). However, these objects remain poorly characterised at older ages (>5 Myr), limiting our ability to constrain the predictions of evolutionary models \citep[e.g.,][]{Somigliana2024,Zagaria2023}. 

With an estimated age range of $\sim$ 5 - 10 Myr \citep{LuhmanEsplin2020}, Upper Scorpius (hereafter USco) is a prime candidate to fill this gap. The first ALMA surveys of the region revealed that disc dust masses were lower than in younger star-forming regions \citep{Carpenter2014, Barenfeld2016}. The spectroscopic surveys of the region available to date either lack access to the most direct tracer of accretion, UV-excess emission \citep[e.g.,][]{Prebisch2002,Rizzuto2015,Luhman2012,Fang2023, Delfini2025}, or are limited to small and incomplete sample sizes \citep{Manara2020}. Nonetheless they show that the stellar, disc, accretion and wind properties have evolved by the age of USco, making larger follow-up campaigns essential. 

\citet{Manara2020}  - hereafter M20, based the target selection of their pilot study on the ALMA observations of \cite{Barenfeld2016}, who targeted discs in the region identified from their infrared excesses. The M20 sample was highly incomplete as the objects were confined to two disc dust mass bins with the specific goal of measuring the spread of the \macc-\mdisc\ relation, a key tracer of disc evolution observed in younger star-forming regions \citep{Manara2016b,Mulders2017,Lodato2017,Zallio2024}. In M20, the authors derived these masses from millimetre continuum flux densities, while X-Shooter spectra allowed them to measure stellar and accretion properties. They show that the accretion and disc properties are more evolved than in younger star-forming regions (e.g. Lupus, Chamaeleon I). However, a number of objects still exhibit high accretion rates at this age. The authors also find a large spread in the values of \macc\, at a given \mdisc\. This result is in tension with the expectations of viscous evolution theory \citep{Lodato2017,Rosotti2017} and suggests that other factors, possibly including external photoevaporation, binarity, and dust evolution, may be relevant in USco. The follow-up studies by \cite{Sellek2020} and \cite{Zagaria2022} suggest a similar conclusion. 

More recently, the ALMA Survey of Gas Evolution of Protoplanetary Disks (AGE-PRO) investigated the region \citep{Zhang2025}. \cite{Anania2025} demonstrate that the decrease in the measured gas radii of the discs in USco can be explained, in part, by the moderate level of external photoevaporation present in the region. Population synthesis models by \cite{Tabone2025} suggest that MHD wind-driven models better reproduce the disc fraction, mass, and accretion rate in USco as well as in Ophiuchus and Lupus. In contrast, \citet{Zagaria2023} and \citet{Somigliana2024} show that USco cannot be treated simply as an older counterpart to Lupus or Chamaeleon I and that any comparison must take this into account. However, in order to discriminate between the various models and address open questions about the region, observations of a large sample of discs in USco are essential. Such a sample will play an important role in addressing broader questions on the late stages of disc evolution.

Following Gaia Early Data Release 3 (EDR3) \citep{Gaia2016,GaiaDR3} the population of stars in USco increased significantly \citep{LuhmanEsplin2020}. This led to ALMA observations of 284 objects showing mid-infrared (MIR) excesses \citep{Carpenter2025}, substantially increasing the disc sample. Their millimetre flux densities were used to derive disc sizes and masses for a number of these targets \citep[][;Zagaria et al. in prep]{Zallio2025, Pinilla2025}. We present for the first time, X-Shooter spectra for 88 of these targets. 

In this work we build upon the survey presented by M20 to obtain the stellar and accretion properties for a sample of 127 disc-bearing stars in USco. We selected objects from previous lists of USco members \citep{Barenfeld2016, LuhmanEsplin2020, Carpenter2025}. We combined X-Shooter with ALMA millimetre flux densities to derive the key stellar, accretion, and disc properties for a statistically significant sample of objects in the region. We aimed to describe the bulk properties of the USco region and investigate their relations within the context of their late evolutionary stage. With this larger sample, we provide a statistically robust description of the \macc - \mdisc ~relation in the region. 

The paper is structured as follows. Section \ref{sec: sample} presents the sample, the X-Shooter observations, and the data reduction. In Sect. \ref{sec: method} we present the analysis of the spectra and the derivation of the stellar and accretion properties. We present the results in \ref{sec: results}. We discuss the findings and their implications in section \ref{sec: discussion} before summarising and concluding in Sect. \ref{sec: conclusions}.

\section{Sample, observations, and data reduction}\label{sec: sample}

\subsection{Sample}
In this survey we build upon the work of M20 who studied 36 target. By adding an additional 91 objects, we more than tripled the sample of disc-bearing stars in USco with measurable disc and accretion properties, to a total of 127. We included objects in a wider range of stellar and disc-mass bins, allowing for a more robust statistical analysis of the region. 

We based our X-Shooter sample selection on the work of \citet{LuhmanEsplin2020}, who classified stars with evidence of discs from near- and MIR excesses and identify them as likely members of USco. Their sample is included in the list of 284 objects recently observed by \citet{Carpenter2025} and, in most cases, shows detected continuum emission with ALMA Band 7 (0.8-1.1 mm). Here we present new X-Shooter observations for 91 objects from their sample that still lack broadband spectra. We include the initial 36 targets of M20 for further completeness.

The targets range in spectral type (SpT) from M6 to G9. From the full sample of 127 targets, we derived stellar and accretion parameters for 121. For the remaining six targets\footnote{The targets with no reported fits are 2MASS J16060215-2003142, 2MASS J16064102-2455489, 2MASS J16062383-1807183, 2MASS J16183317-2517504, 2MASS J16111742-1918285, and 2MASS J16083319-2015549}, the poor-quality X-Shooter spectra prevented us from simultaneously and accurately fitting both spectral type and accretion luminosity (see Sect. \ref{sec: method}). We therefore exclude these targets from the analysis. From the initial M20 sample, we re-analysed the X-Shooter spectra of 35 targets for self-consistency. In one case (2MASS J16042165-2130284) we adopted the M20 parameter values, as the new version of the fitting programme (Sect. \ref{sec: method}) also did not converge. 

The study of irregular dippers in USco by \cite{Empey2025} includes seven objects also present in our sample\footnote{The dippers in our sample are 2MASS J16041893-2430392, 2MASS J16042165-2130284, 2MASS J15583692-2257153, 2MASS J16020757-2257467, 2MASS J16090075-1908526, 2MASS J16141107-2305362, and 2MASS J16154416-1921171}. \citet{Empey2025} show that the observed variability is a result of occultations of the stars from the disc, rather than from sudden bursts of accretion. Hence, we include them in our sample. The total sample considered in the analysis contains 121 targets.

\subsection{Observations}
The observations presented here were obtained with the X-Shooter spectrograph, mounted on ESO's VLT \citep{vernet2011}. The initial 36 M20 targets were observed in service mode under Pr.Id 097.C-0378 (PI Manara) and in visitor under programme Pr.Id. 0101.C-0866 (PI Manara). The remaining 88 targets were observed in service mode programmes, Pr.Ids 105.2082.003, 111.255B.001 (PI Claes, May 2023), 113.26NN.001, 113.26NN.003 (PI Manara, May-September 2024), and P115.27XL.001 (PI Manara, May-September 2025). For all targets, spectra were obtained using both narrow and wide slit widths. The former were taken in an ABBA nodding pattern to better remove sky emission, while the latter were taken in stare mode for the most time-efficient observations. For narrow-slit observations, slit widths were set to 1.0" or 0.5" in the ultraviolet-blue (UVB) arm, and 0.4" or 0.9" for the visual (VIS) and near-infrared (NIR) arms (see Appendix Tab. \ref{tab: logobs}). These configurations ensured spectral resolutions of $\gtrsim$ 10,000 in the VIS and NIR arms ($\lambda >$ 500 nm) and $\gtrsim$ 5500 in the UVB arm ($\lambda \sim$ 300 - 500 nm). Wide-slit observations used a slit width of 5.0" in all arms to ensure maximum flux gain throughout. These were used to correct the higher-resolution, narrow configuration observations for slit losses. 

The observing conditions were generally good for most observing runs, with mostly clear or photospheric conditions and typical airmass-corrected seeing of $\sim 1.0$" in the VIS. For a few cases where the conditions did not meet the requirements, we obtained repeat observations. For Pr.Id 105.2082.003, the observations of four targets were taken in thin and thick conditions with seeing $\sim 2.0$". In visual binary systems, the slits were aligned to include both components, but in general the slits were orientated at a parallactic angle. We present the log of the observations in Appendix. \ref{sec: logobs}.

\subsection{Data reduction}\label{sec: data_reduction}

We carried out standard data reduction procedures using the X-Shooter pipeline v3.6.8 \citep{modigliani2010} alongside the EsoReflex workflow v2.11.5 \citep{freudling2013}. This performs the usual steps of flat, bias, and dark correction, wavelength calibration, spectral rectification, and extraction of the 1D spectra. We completed the flux calibration using a standard star observed on the same night. We performed telluric correction on all spectra using Molecfit for the VIS and NIR arms \citep{Smette2015, Kausch2015}. To correct for slit losses, we scaled the narrow-slit spectra to the wide-slit flux, as in other works \citep[e.g. M20, ][]{Manara2021}. The overall flux calibration is in good agreement with historical photometry for the majority of targets\footnote{Due to the poor observing conditions for 4 targets in Pr.Id 105.2082.003, flux calibration was unreliable and therefore the spectra were scaled to match the historical photometry.}. The seven targets showing significant differences exhibit evidence of ongoing variability (see Appendix~\ref{sec: vars}). 

For several binary targets in the sample (2MASS J16064102-2455489, 2MASS J16060215-2003142, 2MASS J16054540-2023088, and 2MASS J15354856-2958551) we resolved both components. In these cases, we manually extracted the narrow- and wide-slit exposures of both components from the reduced 2D slit images using the Image Reduction and Analysis Facility (IRAF) \citep{IRAF1986,IRAF1993,IRAF2025}. We then flux calibrated them individually. The final spectra of the fainter secondary components often have poor signal-to-noise ratios (S/N). We analysed only the spectra of the primary components, with the exception of 2MASS J15354856-2958551 (east and west) where both components show clear spectra and were identified as young stellar objects (YSOs, see M20). Additionally, due to the poor observing conditions, we were unable to resolve the two components of 2MASS J16113134-1838259 (AS 205). Hence, we excluded this target from our sample.

\section{Data analysis}\label{sec: method}

\subsection{Derivation of stellar and accretion properties}\label{sec: hrd}
We derived the stellar and accretion properties of the targets in the sample by analysing their spectra with the FitteR for Accretion ProPErties of T Tauri stars (FRAPPE \footnote{FRAPPE GitHub: \url{https://github.com/RikClaes/FRAPPE.git}}) code described by \cite{Claes2024}, based on the original method of \cite{Manara2013_fitter}. The FRAPPE code uses an interpolated grid of class III templates to derive spectral properties with an improved fitting method. 

The code reproduces an input spectrum by using three main components : (a) continuum slab models to describe the excess continuum emission from the accretion shock, (b) a reddening law to account for extinction, and (c) a grid of interpolated class III templates to reproduce the stellar photospheric and chromospheric emission. 

The isothermal hydrogen slab models developed in \cite{Rigliaco2012}, \cite{Manara2013a}, and described in \citet{Manara2014PhD}, are used to reproduce the excess UV emission in spectra due to accretion. The models are matched to the de-reddened observed spectra through a scaling factor that is used in the calculation of the total accretion flux.  
The extinction is determined by incorporating the Cardelli reddening law \citep{Cardelli89}. The total-to-selective extinction ratio is fixed to the commonly accepted interstellar value of $R_{\rm V} = 3.1$, while exploring a range of input extinction magnitudes ($A_{\rm V}$) to determine the best fit. The stellar contribution is found by comparing the de-reddened observed spectrum with a grid of interpolated class III templates (see \citealt{Claes2024} for details) ranging from SpT G8 to M9.5. The code applies a scaling factor to the template to correct for distance and stellar luminosity, and aims to reproduce the observed flux at key regions of the spectra. Under this approach, the spectral type determination is typically accurate to within half of a subclass; however, the uncertainties can be larger at the edge of the grid (i.e. for earlier SpTs). 

\begin{figure}
    \centering
    \includegraphics[width=0.95\linewidth]{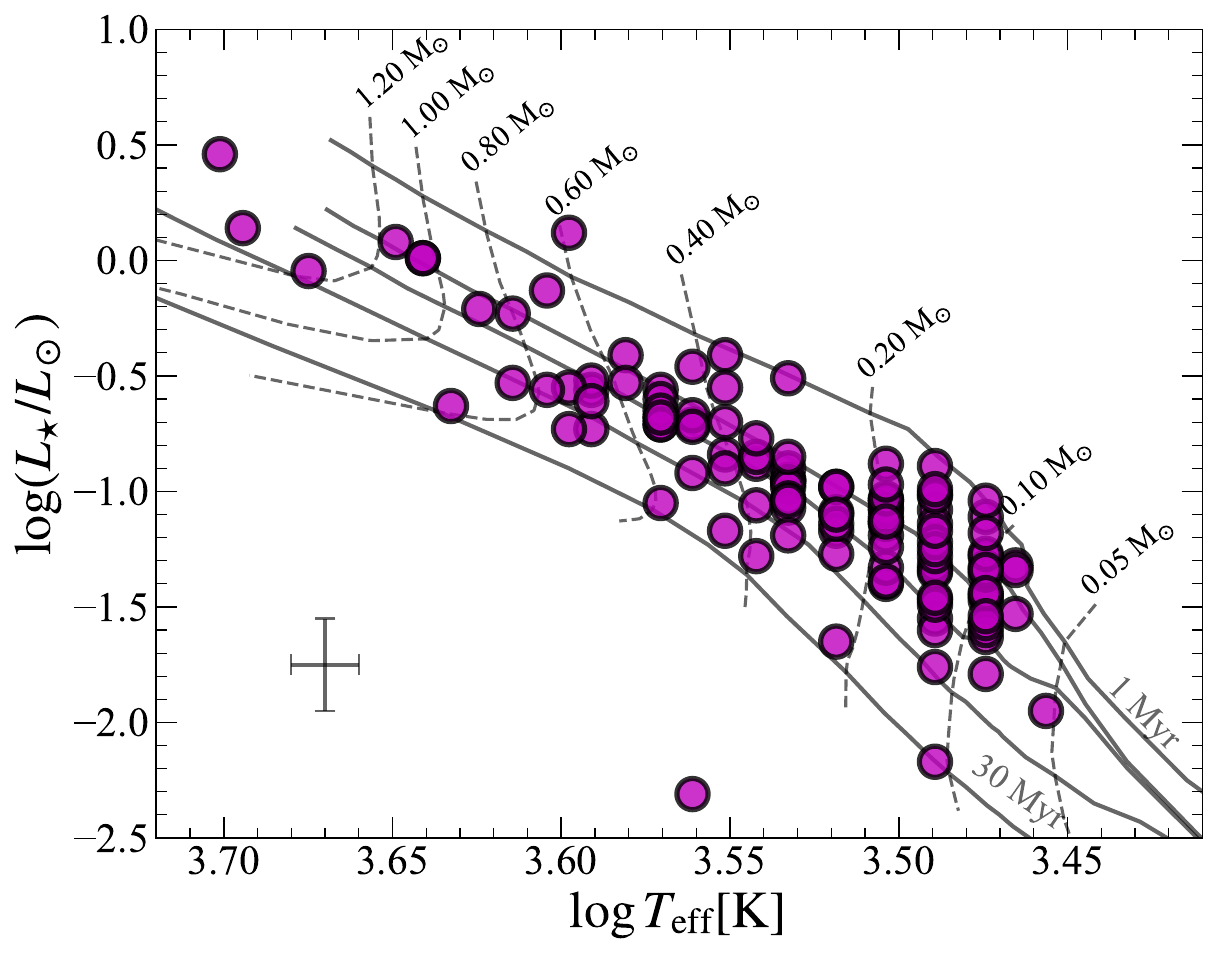}
    \caption{Hertzsprung-Russel (HR) diagram for the sample of disc-bearing stars in USco. The black lines show the \cite{Baraffe2015} evolutionary tracks (dashed) with isochrones (solid), corresponding to ages of 1, 3, 5, 10, and 30 Myr. The grey cross in the lower-left corner represents the typical uncertainties of \lstar\ (0.2 dex) and of \teff\ (0.01 dex). }
    \label{fig: hr_diagram}
\end{figure}

The fitter is ran with a set of input spectral types, range of extinction magnitudes ($A_{\rm V}$) and on the whole grid of slab models, and determines the best fit by minimizing a $\chi^2_{\rm like}$ likelihood function. The SpT is used to obtain the effective temperature ($\rm {T_{eff}}$) via the relations of \cite{HerczegHillenbrand2014}. The stellar luminosity (\lstar) is then computed from $L_\star = 4 \pi d^2 F_{\rm bol}$, where $d$ is the distance to the target and $F_{\rm bol}$ is the stellar flux at 751 nm with the bolometric correction of \citet{HerczegHillenbrand2014} \citep{Claes2024}. By placing these measurements on the HR diagram the stellar age and mass are derived by interpolating between evolutionary tracks. We report the values derived from the tracks of \cite{Baraffe2015}, with the exception of one target (2MASS J15583692-2257153, highest $T_{\rm eff}$ in Fig. \ref{fig: hr_diagram}) for which we use those of \cite{Seiss2000} as it lies outside the parameter range of the Baraffe tracks. 

The accretion luminosity is computed from the slab model flux ($F_{\rm acc}$) integrated over frequency ($L_{\rm acc} = 4\pi d^2 F_{\rm acc}$). The corresponding mass accretion rate is then given by $\dot{M}_{\rm acc} = 1.25 L_{\rm acc} R_\star/(GM_\star)$ \citep{Gullbring1998}. We report the derived properties from FRAPPE for all objects in the sample in Appendix Tab. \ref{tab: master_props} (see \citet{Claes2024} for more details). As the fitter does not report individual uncertainties on the measured values, we include typical uncertainties on the key properties \footnote{Typical uncertainties are as follows : $~\sigma L_\star = 0.2 ~\rm{dex}, ~\sigma M_\star = 0.1 \rm{dex}, ~\sigma L_{\rm acc} = 0.2 ~\rm{dex}, ~\sigma \dot{M}_{\rm acc} = 0.35 \rm{dex} $}. We took parallactic distances ($d$) for each target from the Gaia EDR3 catalogue of \citet{Bailer-Jones2021}.

Figure \ref{fig: hr_diagram} shows the HR diagram of this sample, with one object exhibiting a very low \lstar\ for its associated \teff: 2MASS J16020429-2231468. It is likely that this object was observed in a period of variability (see Appendix \ref{sec: vars}) and so we excluded it from further analysis. The figure also shows that the majority of our targets fall between the 3-10 Myr isochrones of the evolutionary models of \cite{Baraffe2015}. This is generally in line with the expected typical age of the USco region ($\sim$ 5-10 Myr). Seven objects fall below the 10 Myr line, with multiple showing isochronal ages up to 31 Myr. The median isochronal age of the sample is 3.74 Myr; however, this value depends on the choice of evolutionary model \citep[][]{FangHerczeg2025,Zallio2026}. Dynamical mass measurements can also be used to help constrain disc ages \citep{Towner2026,Zallio2026}. However, we did not use individual target ages in this analysis.

\subsection{Upper limits and chromospheric emission}\label{sec: uplims}

Across the sample we find a significant fraction of objects with notably low accretion luminosities (i.e. $L_{\rm{acc}}\lesssim 10^{-4} L_\odot$). At this level, measuring a true accretion luminosity (due to a continuum excess) may be complicated by emission from the stellar chromosphere or by low S/N in the spectra at $\lambda<$ 360 nm. Although FRAPPE can fit these targets, the current version does not estimate individual uncertainties on its \lacc\ measurement \citep{Claes2024}. Therefore, we introduced a criterion to separate definite accretors from targets for which the excess is uncertain, and for the latter we estimated 1 $\sigma$ upper limits of the accretion measurements. In brief, the method distinguishes objects based on the S/N of the observed spectra in the four bins of the Balmer continuum used by FRAPPE to determine the excess emission due to accretion. We provide more details of this method in Appendix \ref{sec: uplims_details}. Using this approach, we classify a total of 49/121 (40\%) objects as upper-limit measurements.

\begin{figure}
    \centering
    \includegraphics[width=0.95\linewidth]{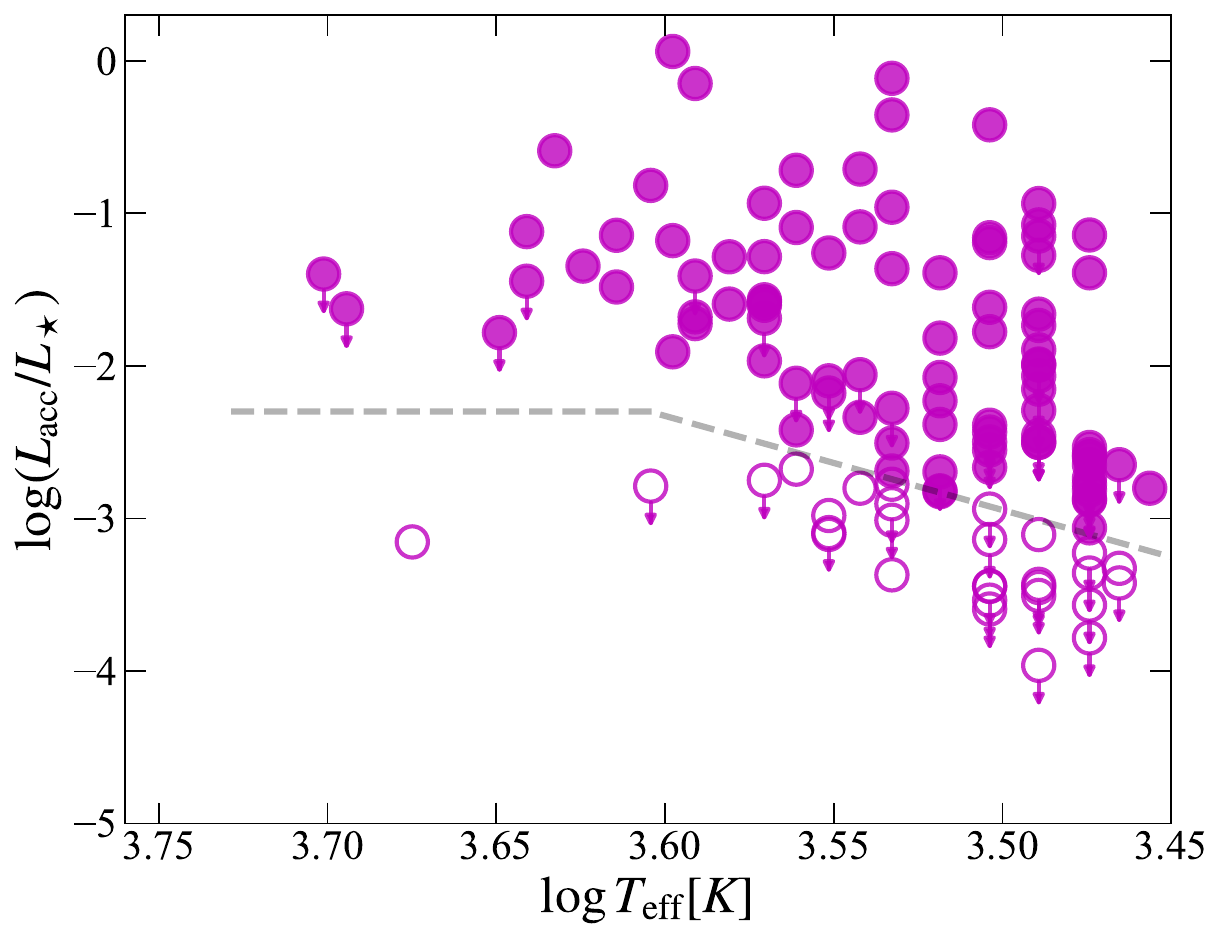}
    \caption{Accretion-to-stellar luminosity (\lacc/\lstar), as a function of effective temperature (\teff). The dashed grey line shows the chromospheric emission limit derived by \cite{Manara2013a}. Objects having a ratio equal to or below this line may be contaminated by chromospheric emission and are shown with hollow markers. Upper limits (Sect. \ref{sec: uplims}) are indicated with downward-facing arrows.}
    
    \label{fig: chromo_limit}
\end{figure}
  
T Tauri stars are known to have active chromospheres \citep{Houdebine1996, Franchini1998, Stelzer2013}, whose emission can contaminate measurements of accretion luminosity \citep{Inglby2011,Rigliaco2012}. To investigate this effect, we compared the \lacc/\lstar\ ratio of our objects to typical chromospheric emission levels of class III stars of the same spectral type \citep[][ see Fig. \ref{fig: chromo_limit}]{Manara2013a,Manara2017}. This comparison provides a simple check to flag objects whose \lacc~measurements may be affected by chromospheric emission. We find that 29 targets in the sample lie on or below this threshold. These targets predominately populate the lowest measured accretion luminosity range ($L_{\rm acc} \lesssim 10^{-4} L_\odot$). Therefore, any measured continuum excess may not be due solely to ongoing accretion, but may be contaminated by chromospheric emission \citep{Manara2017}. Previous works have considered such targets to be possible non-accretors \citep{Alcala2014, Alcala2017, Manara2016b, Manara2017,Manara2020,AlmendrosAbad2024}.

Of the 29 objects whose ratios fall below the chromospheric emission limit, 19 also have upper-limit accretion measurements from their low Balmer continuum excess (S/N). The remaining ten form an interesting sub-sample with well-measured continuum excesses that are, however, below the chromospheric limit for their spectral type. This is possible if the photospheric template used in the analysis has lower chromospheric noise than typical YSOs or the target star has an anomalously high level of chromospheric activity. 

For simplicity we henceforth categorise the objects in our sample into two groups: definite accretors and upper-limit accretors. There is an  approximately even split between these two categories, with 62 and 59 targets, respectively. The former are characterised by a clear UV continuum excess and have accretion luminosities above the chromospheric activity limit for their SpT (filled markers and no arrows in Fig. \ref{fig: chromo_limit}). We classify the remaining objects upper-limit accretors; this can occur because their spectra show no excess with respect to the interpolated Class III template, poor S/N in the UV, or because the measured accretion luminosity falls below the chromospheric activity threshold. In the following sections they are treated as censored data in all statistical analysis.

\subsection{Complementary ALMA data}

We computed the disc dust masses of the entire sample from the ALMA Band 7 continuum flux densities reported by \citet{Carpenter2025}. The resulting values are reported in Table \ref{tab: master_props}. As in previous studies, and mindful of the many caveats \citep[e.g.][]{Miotello2023}, we used this information as a tracer of disc dust mass. To convert the flux densities to dust masses, we assumed isothermal, optically thin dust emission from the disc following the approach of \cite{Barenfeld2016}. Using distances from Gaia EDR3 (as mentioned in Sect. \ref{sec: method}) we estimated the dust temperature profile with a Planck function with characteristic temperature, $T_d$. In this study we adopted the stellar luminosity scaling relation $T_d = 25 K \times(L_\star / L_\odot )^{0.25}$. We used the \lstar values derived in this work. We note that our results do not change if we assume a fixed temperature of $T_d = 20 K$, as is done in other studies. We computed total disc masses, hereafter referred to as $M_{\rm disc}$ for simplicity, assuming the standard 1:100 dust-to-gas ratio (i.e. $M_{\rm disc} = 100 \cdot M_{\rm dust}$). This ratio is commonly assumed in the literature, but assumes little evolution with time in discs. We discuss this aspect in Sect.~\ref{sec: discussion}. Four targets\footnote{The targets with unreliable millimetre flux density measurements are: 2MASS J16185382-2053182, 2MASS J16120505-2043404, 2MASS J16064385-1908056, and 2MASS J15442550-2126408} were flagged as having unreliable millimetre flux densities from ALMA; as a result we did not include them in the analysis of disc masses.

In addition, Zagaria et al. (in prep.) derive CO disc radii for a total of 90 discs from the ALMA data of\cite{Carpenter2025}. We include the median, 16th percentile and 84th percentile values of the 90\% CO flux radii for 73 targets in our sample (see Sect. \ref{sec: results}). We also include the disc structure classifications derived from the ALMA observations of \cite{Carpenter2025, Pinilla2025, Vioque2025} for the corresponding targets in our sample.

\begin{table}
\caption{Results of the power-law fits to relations between stellar, disc and accretion properties.} 
\centering
\begin{tabular}{@{} c | c c c c @{}}
\hline
\hline
\textbf{Relation}        &  \multicolumn{4}{c}{\textbf{Upper Scorpius best fit parameters}} \\ 
        y-x        &$\alpha$ & $\beta$ & $\sigma$ & $\rho$\\ \hline
\lacc - \lstar &   $-1.6 \pm 0.4$  & $2.1\pm0.4$ & $1.4 \pm 0.7$ & $0.5 \pm 0.1$\\ \hline
\macc - \mstar  &  $-9.1 \pm 0.3$ & $2.4 \pm 0.5$ & $1.3 \pm 0.6$ & $0.5 \pm 0.1$ \\ \hline
\macc - \mdisc  &  $-6.9 \pm 0.9$ & $1.1 \pm 0.3$ & $1.5 \pm 0.7$ & $0.4 \pm 0.1$ \\ \hline
\macc - $R_{\rm{CO}}$  &   $-10.4 \pm 1.4$ & $0.0 \pm 0.8$ & $1.5 \pm 0.8$ & $0.0 \pm 0.1 $ \\ \hline
$M_{\rm dust}$ - \mstar &  $0.7 \pm 0.1$  & $0.8 \pm 0.1$ & $0.5 \pm 0.2$ & $0.4 \pm 0.1$\\ 
\hline
\end{tabular}
\tablefoot{The relations are always in the form $\log_{10} {\rm y} = \alpha + \beta \log_{10} {\rm x}$. 
$\sigma$ measures the dispersion from the fit, $\rho$~is the correlation coefficient where a value of 1 represents a perfect correlation and 0 signifies no correlation (see \textit{linmix} docs). The fit of each relation is shown in Figures \ref{fig: lacc_lstar},\ref{fig: macc_mstar}, \ref{fig: macc_mdisc}, \ref{fig: macc_Rco} and Appendix fig. \ref{fig: mdust_mstar_linmix}. The fits are based on the values reported in Table \ref{tab: master_props}, for which stellar properties were derived using the evolutionary tracks of \citet{Seiss2000, Baraffe2015}. }
\label{tab: linmix_params}

\end{table}

\section{Results}\label{sec: results}

This section summarises the derived stellar, disc, and accretion properties of the sample before exploring several key relations (\lacc-\lstar, \macc-\mstar, \macc-\mdisc, and \macc-\Rco) for the disc-bearing objects in USco. We report the measurements of the stellar, accretion and disc properties for the sample in Appendix Table~\ref{tab: master_props}. The large number of upper-limit measurements makes it difficult to accurately determine statistically meaningful medians and quantiles for the measured properties (see Appendix \ref{sec: add_plots}). We therefore focused our analysis primarily on the observed scatter and on fitting the relations with a power law, as was done in similar studies of other star-forming regions. We determined the best-fit power law using the \textit{linmix} Python package, which implements the hierarchal Bayesian linear regression method described by \citet{Kelly2007}. This approach allowed us to include upper limits in the fitting procedure. Table \ref{tab: linmix_params} displays the best fit parameters.

\subsection{Accretion-stellar luminosity relation}
\begin{figure}
    \centering
    \includegraphics[width=0.95\linewidth]{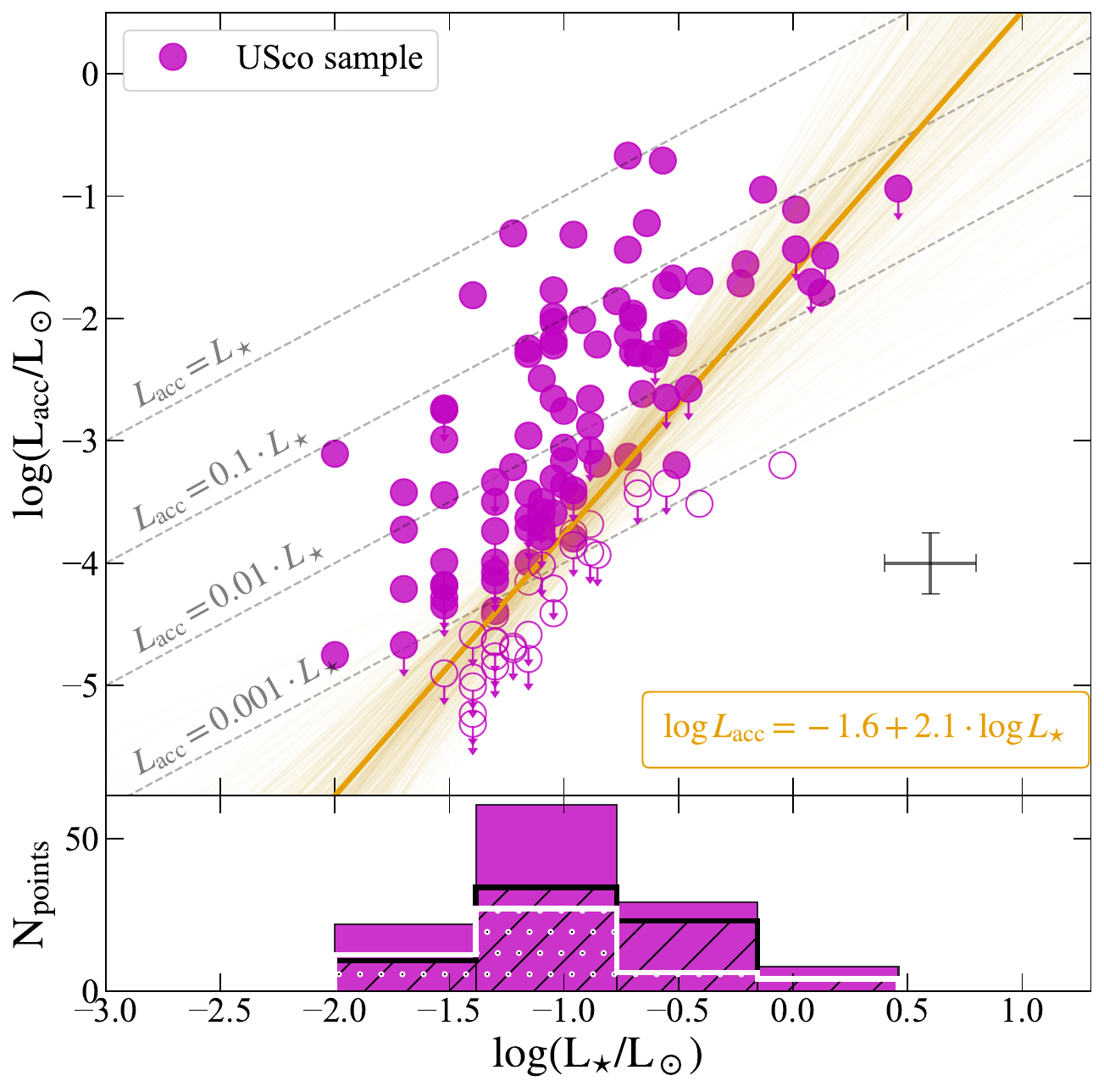}
    \caption{Top panel: Accretion luminosity (\lacc) as a function of stellar luminosity (\lstar). Solid magenta points are definite accretors, and those with downward-pointing arrows indicate upper-limit measurements based on their continuum excess. Hollow magenta points indicate targets with $L_{\rm acc}/L_\star$ ratios below the chromospheric limit (see Fig. \ref{fig: chromo_limit}). Overplotted in orange is the best fitting power law shown in the textbox (determined by \textit{linmix}, see Table \ref{tab: linmix_params} for full fit parameters and uncertainties). Faded coloured lines show sample fits taken from the Bayesian posterior \citep{Kelly2007}. Dashed grey lines show different fixed dependencies of \lacc\ on \lstar to guide the eye. The grey cross represents typical uncertainties in \lacc, and \lstar. Bottom panel: Histograms of the full sample (magenta), upper-limit accretors (white dotted), and definite accretors (black striped) in each \lstar\, bin. }
    \label{fig: lacc_lstar}
\end{figure}

Fig. \ref{fig: lacc_lstar} shows the distribution of accretion luminosity as a function of stellar luminosity. The plot shows measurements of definite accretors as solid magenta points, and those of upper-limit accretors with arrows and/or hollow points. Hollow points represent objects whose \lacc / \lstar ratio falls below the expected chromospheric activity limit. The histogram shows that the proportion of upper limits to definite detections is approximately 50:50 in all bins. The upper-limit accretors are distributed across the full \lstar\ range. As expected, the majority are among the lowest values of accretion luminosity in the sample. In contrast, several objects show very high accretion luminosities (\lacc $\sim L_\odot$). This demonstrates the wide range of accretion levels among the targets in the sample. In the central \lstar\, bin ($-1.3 \le \log (L_{\star}/L_\odot) \le -0.7$), we observe a large vertical spread in \lacc, that spans approximately four orders of magnitude. In general, with the exception of the highest \lstar values, \lacc\ shows a significant spread across all bins. In Fig. \ref{fig: lacc_lstar}, we also observe a cut-off at around $\log (L_\star/L_\odot) \approx 0$, indicating a lack of objects with solar luminosity or higher.

To further evaluate this relation we fitted the scatter with a power-law relation (solid orange line in Fig. \ref{fig: lacc_lstar}). The effect of the upper-limit values on the fit is evident from the best-fit curve passing through the lower envelope of the scatter. We measure a slope ($\beta$) of $2.1 \pm 0.4$, indicating a relatively strong dependence of the accretion luminosity on the stellar luminosity. However, the fitting procedure returns a correlation coefficient ($\rho$) of $0.5 \pm 0.1$, suggesting a moderate positive relation. The large dispersion ($\sigma$) of $1.4 \pm 0.7$ is also consistent with this interpretation. Given the large spread in \lacc, this is unsurprising. Overall, although there appears to be a relatively strong dependence of the accretion luminosity on stellar luminosity, the combination of a large spread and low correlation coefficient makes this result uncertain.

\subsection{Mass accretion rate-stellar mass relation}
\begin{figure}
    \centering
    \includegraphics[width=0.95\linewidth]{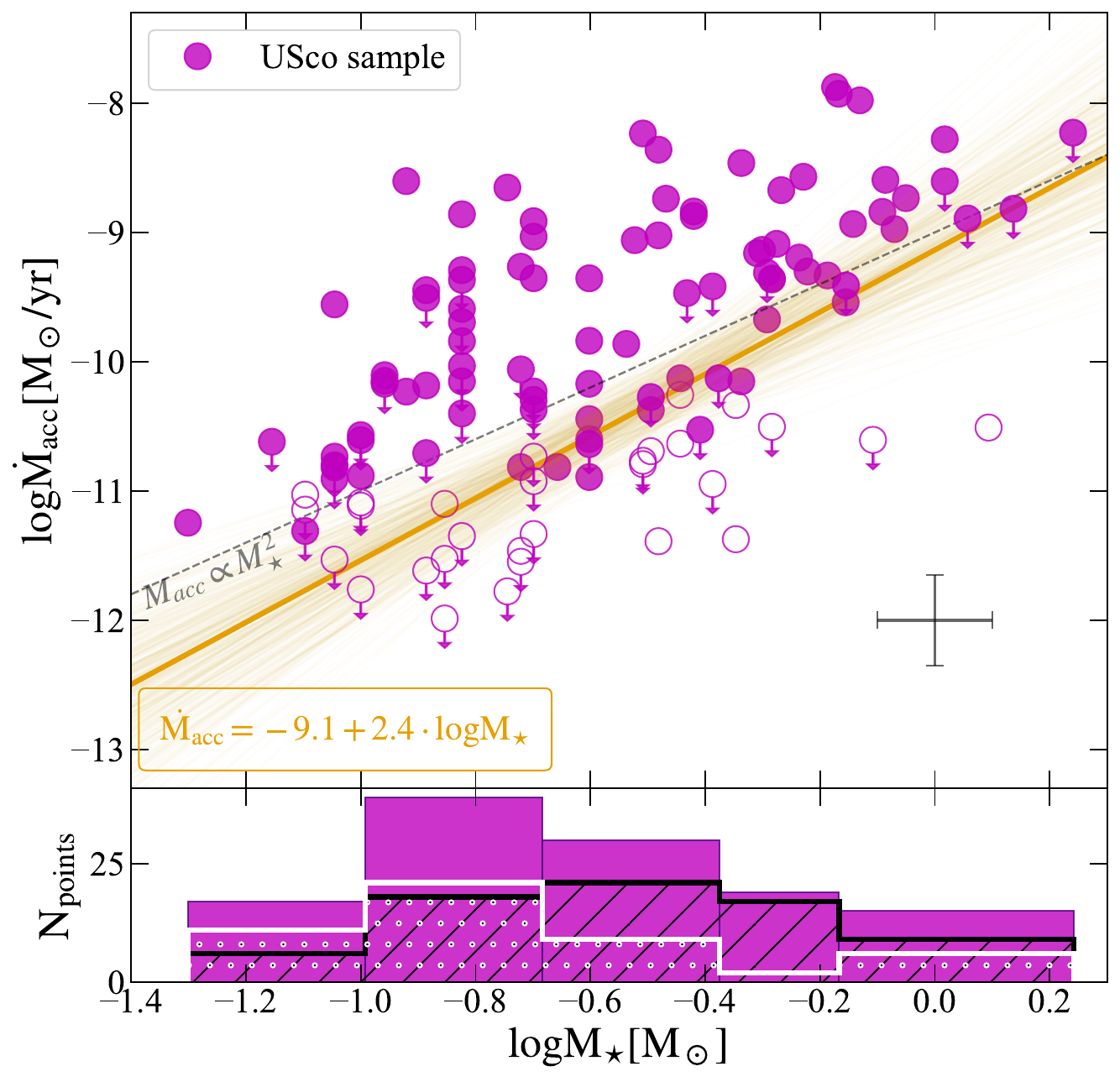}
    \caption{Top panel: Mass accretion rate (\macc) plotted as a function of stellar mass (\mstar). Symbols and power-law fit are as in Fig. \ref{fig: lacc_lstar}. The dashed grey line shows the \macc $\propto M_{\star}^2$ relation to guide the eye. The grey cross represents typical uncertainties in \macc\ and \mstar. Bottom panel: Histograms of the full sample (magenta), upper-limit accretors (white dotted), and definite accretors (black striped) in each \mstar bin.}
    \label{fig: macc_mstar}
\end{figure}

Figure \ref{fig: macc_mstar} shows the mass accretion rate as a function the stellar mass. \footnote{We note that the results for this relation can depend on the adopted evolutionary tracks used to derive the stellar mass \citep[e.g.][]{Betti2023,AlmendrosAbad2024}. As previously mentioned, we used the values reported in Appendix Tab. \ref{tab: master_props}, which are based on the tracks of \citet{Seiss2000, Baraffe2015}.} As highlighted by the histogram at the bottom of the plot, upper limits on \macc\, are present across all \mstar~bins, but their fraction relative to definite detections is higher in the lower \mstar\ bins. We observe a very large spread in \macc\ at all values of \mstar. This spread reaches just over three orders of magnitude in the two central \mstar~ bins. Despite this, the sample shows evidence of an increasing trend in \macc\ with increasing \mstar. 

To quantify this trend and facilitate comparison with results obtained in other star-forming regions, we fitted the data with a power law, using the same approach as for \lacc-\lstar. As expected, the fit is significantly influenced by the large number of upper limits throughout the sample. The fitting procedure returns a best-fit value of $\beta = 2.4 \pm 0.5$, indicating a strong dependence and therefore relation between the two properties. Figure \ref{fig: macc_mstar} also includes a $\dot{M}_{\rm acc} \propto M_\star ^2$ relation for context, as this has been observed in other star-forming regions \citep[e.g.][]{Hartmann2016}. Comparison with our best-fit power law shows little difference. The dispersion of the points around the fit ($\sigma$) is also significant (see Tab. \ref{tab: linmix_params}), with a value of $1.3 \pm 0.6$. Together with a moderate correlation coefficient ($\rho =0.5 \pm 0.1$), this again indicates an uncertain dependence driven by the large spread in \macc. The fit also returns $\alpha = -9.1 \pm 0.3$. 

The measured slopes are consistent with previous values \citep[e.g.][and references therein]{Hartmann2016,Manara2023}, but the correlation is weaker than previously found. Furthermore, the targets in the upper envelope of Fig. \ref{fig: macc_mstar} may also be consistent with a double power law fit, in which \macc\ depends more strongly on lower \mstar\ objects than those with \mstar$\geq 0.2 M_\odot ~(\log M_\star \approx -0.7)$. This behaviour was predicted by \citet{Vorobyov2009}, and is observed in other star-forming regions \citep[e.g.,][]{Alcala2017,Manara2017}. However, given the significant spread in the sample, we did not attempt to fit this here. We further discuss the \macc-\mstar\ relation in comparison with other regions in Section \ref{sec: discussion}.

\subsection{Mass accretion rate-disc mass relation}
\begin{figure}
    \centering
    \includegraphics[width=0.95\linewidth]{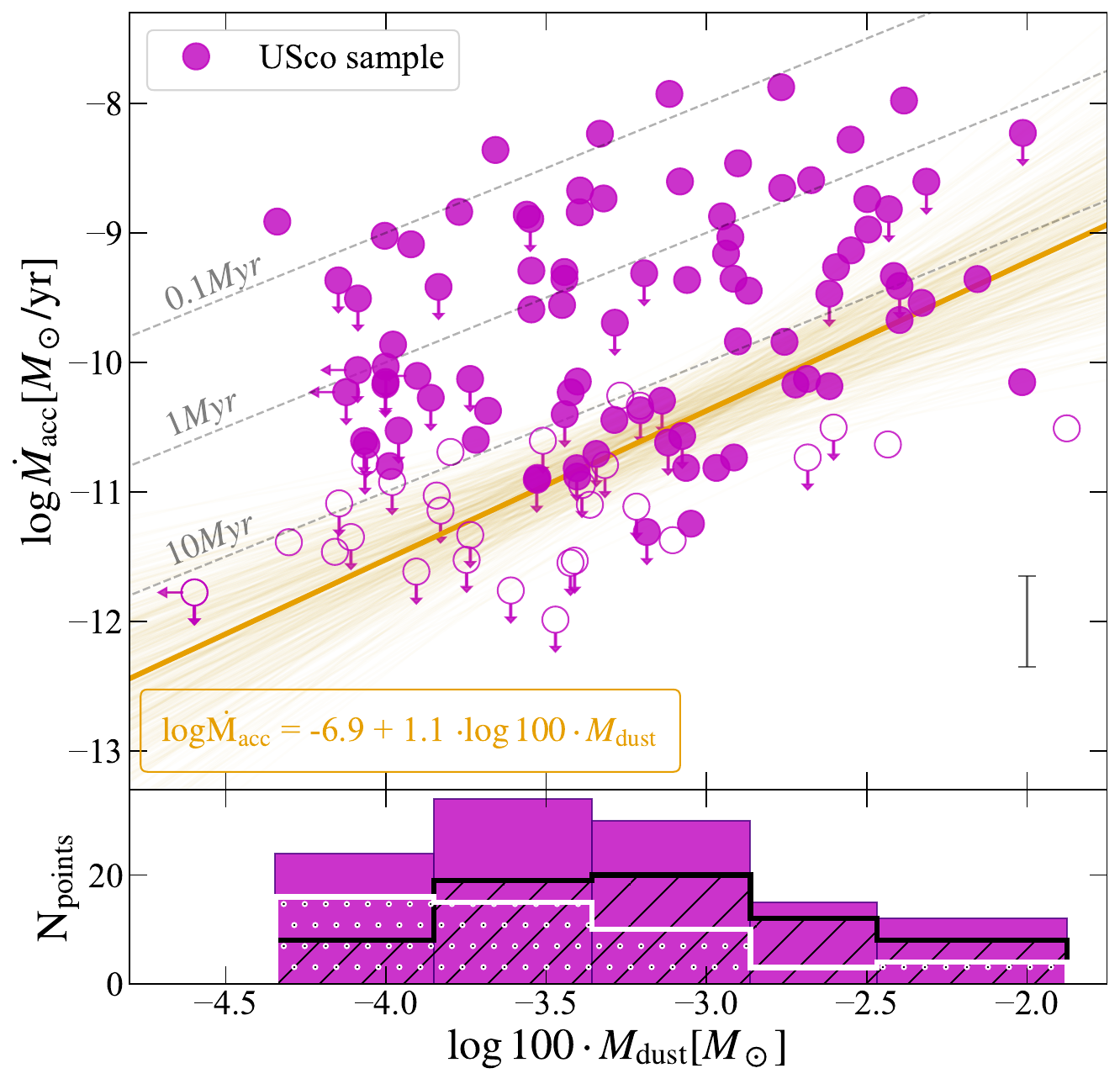}
    \caption{Top panel: Mass accretion rate (\macc) as a function of disc mass (\mdisc). Symbols and power-law fit are as in Fig. \ref{fig: lacc_lstar}. The dashed grey lines indicate \mdisc/\macc~ratios of 0.1, 1, and 10 Myr. The grey error bar indicates the typical uncertainty in \macc. Bottom panel: Histograms of the full sample (magenta), upper-limit accretors (white dotted), and definite accretors (black striped) in each \mdisc~bin. }
    \label{fig: macc_mdisc}
\end{figure}

Figure \ref{fig: macc_mdisc} shows the relation between the mass accretion rate and the disc dust mass. As previously mentioned, we excluded four targets flagged by \citet{Carpenter2025} as having unreliable ALMA millimetre fluxes. We also excluded one object, 2MASS J16160448-2932400, which has only an upper-limit disc dust measurement that is the lowest in the sample ($\log (M_{\rm{disc}}/M_\odot) < -5.0$).

The histogram shows that upper-limit accretors are distributed across the full range of \mdisc, with a higher fraction at the lowest disc masses. As in the previous cases, we also observe a large spread in \macc\ at all values of \mdisc. The spread is roughly consistent across all \mdisc\ bins (approximately 3 dex). Given this spread, there appears to be no relation between the disc dust mass and the accretion rate. The points also do not follow any of the three dashed grey lines representing disc lifetimes ($\tau_{\rm disc} =  M_{\rm disc}/ \dot{M}_{\rm acc}$) 0.1, 1, or 10 Myr. The large spread of the data around these lines at the age of USco is in contrast with the expectations of viscous evolution. We discuss this further in Section \ref{sec: discussion}.

Due to the consistently large scatter across all \mdisc\ values, the best-fit power law shown in Fig. \ref{fig: macc_mdisc} is unconvincing. The line passes through the points just below, and almost parallel to, the 10 Myr $\tau_{\rm disc}$ line, and is once again heavily influenced by the number of upper-limit accretors. It yields a best fit slope of $\beta = 1.1 \pm 0.3$, consistent with previous findings \citep[e.g.][]{Manara2016b,Mulders2017}. However, given the large dispersion ($\sigma = 1.5 \pm 0.7$) and low correlation coefficient ($\rho = 0.4 \pm 0.1$), this indicates a very weak \macc-\mdisc\, dependence.

\subsection{Mass accretion rate-radius relation}
\begin{figure}
    \centering
    \includegraphics[width=0.95\linewidth]{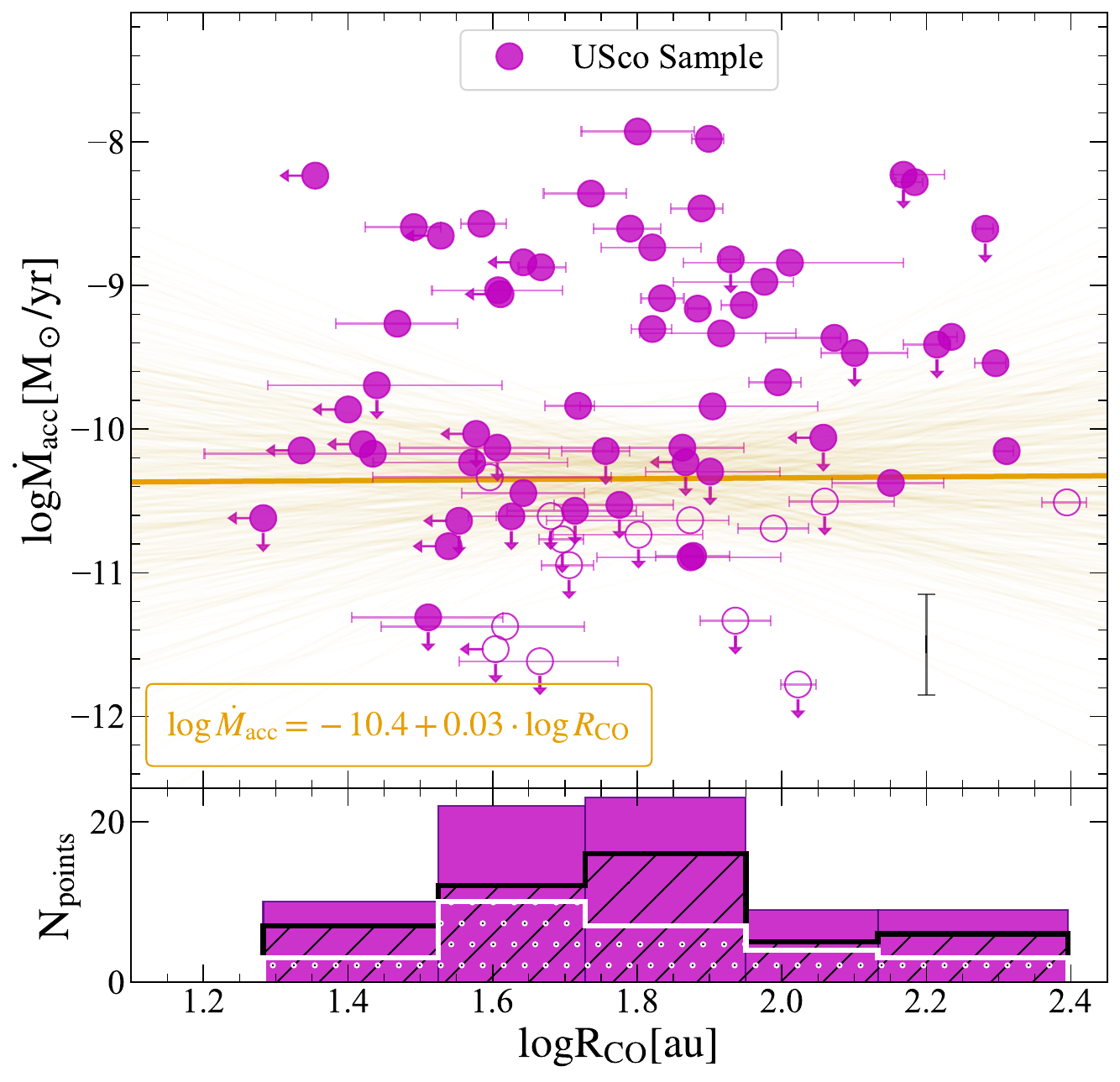}
    \caption{Top panel: Mass accretion rate (\macc) as a function of gas CO disc radius (\Rco). Values of \Rco\ are taken as the median 90\% CO flux radii from Zagaria et al. (in prep), with uncertainties given by the 16th and 84th percentiles of the size distributions. In some cases these are smaller than the size of the points. Symbols and power law fit are as in Fig. \ref{fig: lacc_lstar}. Horizontal and vertical arrows indicate targets for which the reported \Rco or \macc\ value is an upper limit. The grey error bar represents the typical uncertainty in \macc. Bottom panel: Histograms of the full sample (magenta), upper-limit accretors only (white dotted), and definite accretors only (black striped) in each \Rco~bin. }
    \label{fig: macc_Rco}
\end{figure}

Zagaria et al (in prep.) measured gas disc sizes for 73 targets in our sample from the integrated intensity maps of CO (J = 3-2) emission. Figure \ref{fig: macc_Rco} shows the relation between the mass accretion rate and the CO disc radius. As in the previous relations, we observe a significant spread in \macc\, at all disc sizes. This suggests that there is no trend between \macc\ and \Rco. However, the sample contains a few objects with only an upper limit on their CO radii, as well as some with significant error bars. Therefore, it is possible that these errors could mask a correlation between the two properties. Even if this were the case, a large spread in \macc\ values would remain. 

The parameters of the best-fit power law are consistent with this interpretation, given the large dispersion and uncertainties. The slope has a value of $\beta = 0.0 \pm 0.8$, indicating that a weak correlation could be present, although it is more likely that there is none. The correlation coefficient is the lowest in the sample, with a value of $\rho = 0.0 \pm 0.1$, indicating no evidence of correlation even when the uncertainties are taken into account. The absence of a clear relation, together with the spread in \macc~(around four orders of magnitude), suggests that the amount of material accreted onto a star is not related to the size of the disc.

\section{Discussion}\label{sec: discussion}

\subsection{Upper Scorpius membership and subgroups}

Since we selected the X-Shooter sample for this work, the catalogue of disc-bearing stars believed to be members of USco has been updated. Recent studies have relied on further analysis of Gaia data \citep{GaiaDR3} to better constrain the membership of objects to different star-forming regions and even in their subgroups. In particular, \citet{Ratzenbock2023} suggest that 13 objects in our sample are members of $\rho$ Ophiucus (L1688). A further 11 objects were not included in their analysis, and therefore were not linked to any of the USco subgroups. We tested whether their inclusion in our analysis affected our results and find that it did not. First, the 24 potential non-members are distributed similarly 
to the other objects on the HR diagram. Moreover, we recomputed the power law fits for the \macc\ versus \mstar and \macc\ versus \mdisc\ relations presented in Sect. \ref{sec: results} using the reduced sample of 97 bona fide members. The resulting fit parameters are very similar to those listed in Table \ref{tab: linmix_params}, with all differences well within the uncertainties. In particular, the correlation coefficient $\rho$, remains essentially unchanged. 

\citet{Ratzenbock2023} applied their \texttt{SigMA} clustering algorithm to the entire Scorpius-Centaurus OB association, and found that USco can be divided in to 9 distinct subgroups. Using Gaia DR3 data for stars in this region, they identified groups based on their positions and radial velocities. We find that objects in our sample belong to eight of these groups, while the remaining 11 were not included in their study. As discussed above, one of these is $\rho$ Ophiuchus (L1688). The most populous subgroup in USco, $\delta$ Sco, is also the most abundant subgroup in our sample containing a total of 46 objects. The next three most populous subgroups in our sample are $\nu$ Sco, $\rho$ Ophiuchus (L1688), and $\beta$ Sco containing 18, 13, and 12 objects respectively. 

To investigate the effects of subgroup membership on our results we first examined whether the subgroups have different characteristic ages. We calculate the mean isochronal age for each subgroup and find that $\delta$ Sco is the oldest (5.68 Myr), followed by $\nu$ Sco (5.18 Myr), $\beta$ Sco (5.10 Myr), and finally $\rho$ Oph (4.7 Myr). This sequence of ages is broadly consistent with that reported by \citet{Ratzenbock_ages}; the only difference is that they identify $\beta$ Sco as having the second oldest age. However, our measured ages, derived using the \citet{Baraffe2015} evolutionary tracks, are $\sim$ 2.5 Myr younger for two of the three 'Sco' subgroups discussed. We also observe a large spread in ages within each subgroup in our sample, making any measurement of a characteristic age uncertain. 

\begin{figure}
\centering
\includegraphics[width=0.49\textwidth]{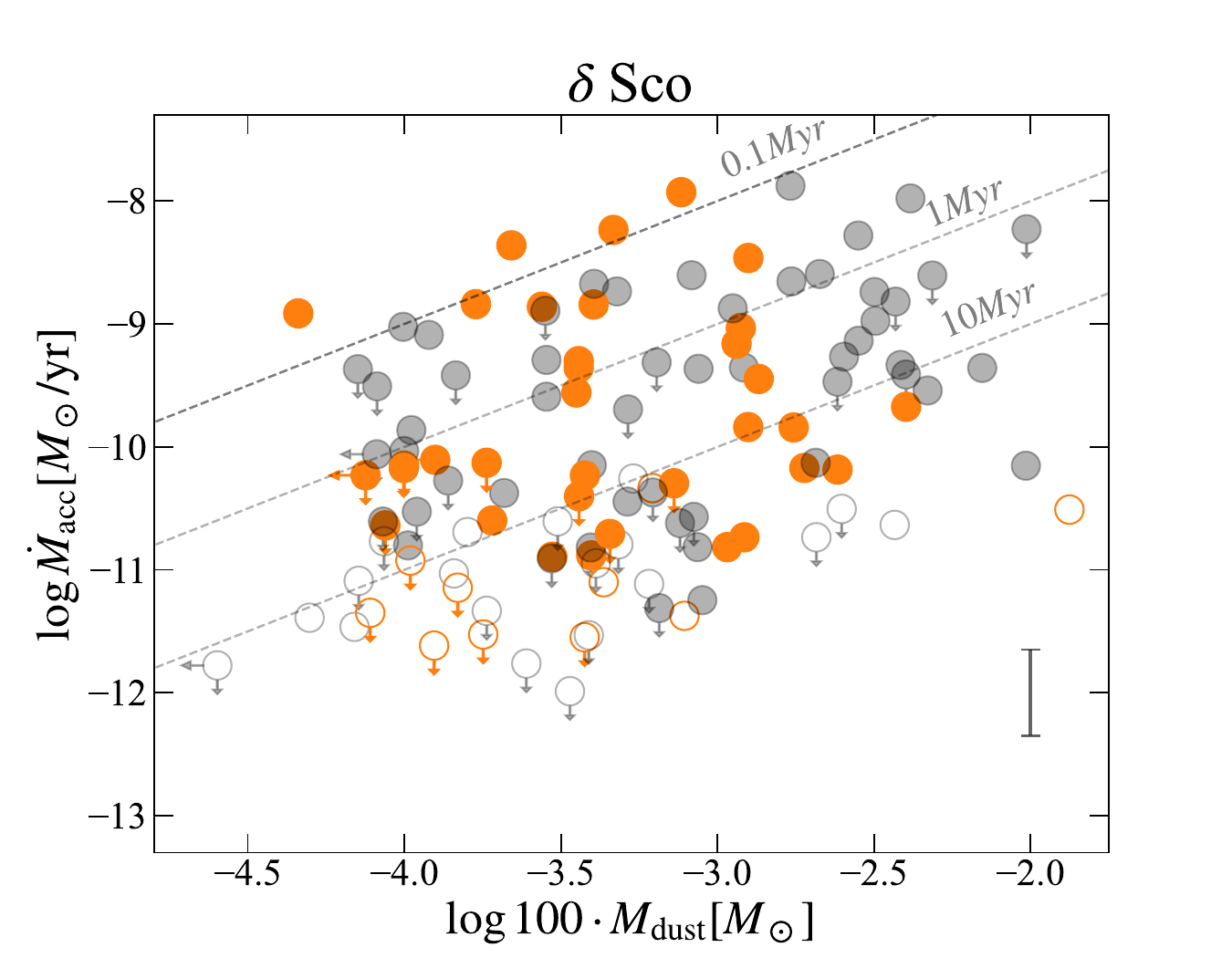}
  \caption{\macc\ as a function of \mdisc for the most abundant USco subgroups in our sample detected with Gaia \citep{Ratzenbock2023}. 
  A total of 112 objects in our sample were are associated with eight subgroups in the region. The sample sizes in each range between 1-45 objects; here we highlight the objects associated with the $\delta$ Sco group (orange points, 45 objects). The remaining objects in our sample are shown in grey. The grey error bar represents the typical uncertainty in \macc.}
     \label{fig: macc_mdisc_deltasco}
\end{figure}

We also investigated the location of objects from each subgroup in the \macc-\mdisc\ plot. For each subgroup, the objects approximately span the full \mdisc~range. Figure \ref{fig: macc_mdisc_deltasco} highlights $\delta$ Sco objects in the \macc-\mdisc\ distribution of our full sample. Although several of the high accretors belong to this group, we find no evidence that subgroup membership explains the large spread observed in our sample. We do observe minor differences in mass accretion rates among the other three example subgroups. The spread in \macc\ may be larger for $\delta$ Sco than for $\nu$ Sco, but the small sample size of the latter makes it is difficult to confirm. If true, this could imply a correlation between the subgroup age and the spread in \macc\ values, potentially mirroring that seen between different star-forming regions. However, drawing any conclusions on the effect of subgroup membership is again limited by the small sample sizes for most subgroups. 

\subsection{The accretion properties of young stars in Upper Scorpius}

In Sect.~\ref{sec: uplims} we assigned upper limits to the accretion measurements for almost half of our sample. This allowed us to classify the sample two groups: definite accretors and upper-limit accretors. Definite accretors show clear evidence of a UV continuum excess due to accretion. upper-limit accretors, however, show no evidence of a UV continuum excess (within the S/N of their spectra) or have measured values below the statistical estimate of the chromospheric activity of the star. For these objects accretion may still occur below what we can measure, but their discs may also have ceased accretion. As in M20, were therefore discuss these objects as potential non-accretors, while acknowledging that may still undergo low levels of accretion.

The separation between definite accretors and potential non-accretors clearly highlights the large dispersion in \lacc\ and, consequently, \macc\ at all stellar masses and disc dust masses. The spread is also not confined to a particular range of stellar or disc properties. The fraction of potential non-accreting targets in this work (approximately half of the sample) is significantly higher than that reported in M20 and previous studies of star-forming regions. This implies that the fraction of accreting stars (51\%) is significantly lower than the fraction of stars showing near-to-mid infrared excess (100\%, as per our selection criteria). This large fraction may be a signature of the late stages of disc evolution. \citet{Fedele2010} found that the fraction of accreting discs decreases with age and is smaller than the fraction of discs that show an infrared excess. They attributed this behaviour to the faster clearing of the inner gaseous disc, traced by accretion, relative to the rest of the dusty disc traced by IR-excess. Furthermore, they estimated an accretion timescale of approximately 2.3 Myr, and found no evidence of ongoing accretion beyond 5 Myr. We derive a longer average accretion time scale ($\tau_{\rm acc} = M_{\rm disc} / \dot{M}_{\rm acc}$) of 5.3 Myr. Given the older age of USco and the number of potential non-accretors in our sample, many of these discs may have ceased accreting, which is what we are seeing. 

Although the sample shows clear signs of ageing, with 49\% of objects having low accretion luminosities a number of objects remain at the opposite end of the scale. Fig. \ref{fig: lacc_lstar} shows that the accretion luminosities span almost five orders of magnitude within a given stellar-mass bin, with the strongest accretors having $L_{\rm acc} \approx L_\star$. Such high luminosities are comparable to those observed in younger star-forming regions, and show that the picture in USco is not so clear. 

In Sect.~\ref{sec:comparison_others}, we show that the highest mass accretion rates in USco are comparable to those of stars with the same stellar or disc mass as those in the younger Chamaeleon I (2.8 Myr) and $\rho$ Ophiuchus (1 Myr) star-forming regions. However, the upper envelope of \macc\ values in our sample, is generally lower than the highest from the other regions. Nonetheless, these high values suggest that there is a real dispersion in the activity of discs in USco. Strong accretors at old ages have also been observed in other regions \citep[e.g.][]{Ingleby2014,Rogers2025} and their existence remains an open question for current theoretical frameworks. We discuss one possible explanation in the next section. 

\subsection {Effect of binarity and disc structures}

\begin{figure}
\centering
\includegraphics[width=0.49\textwidth]{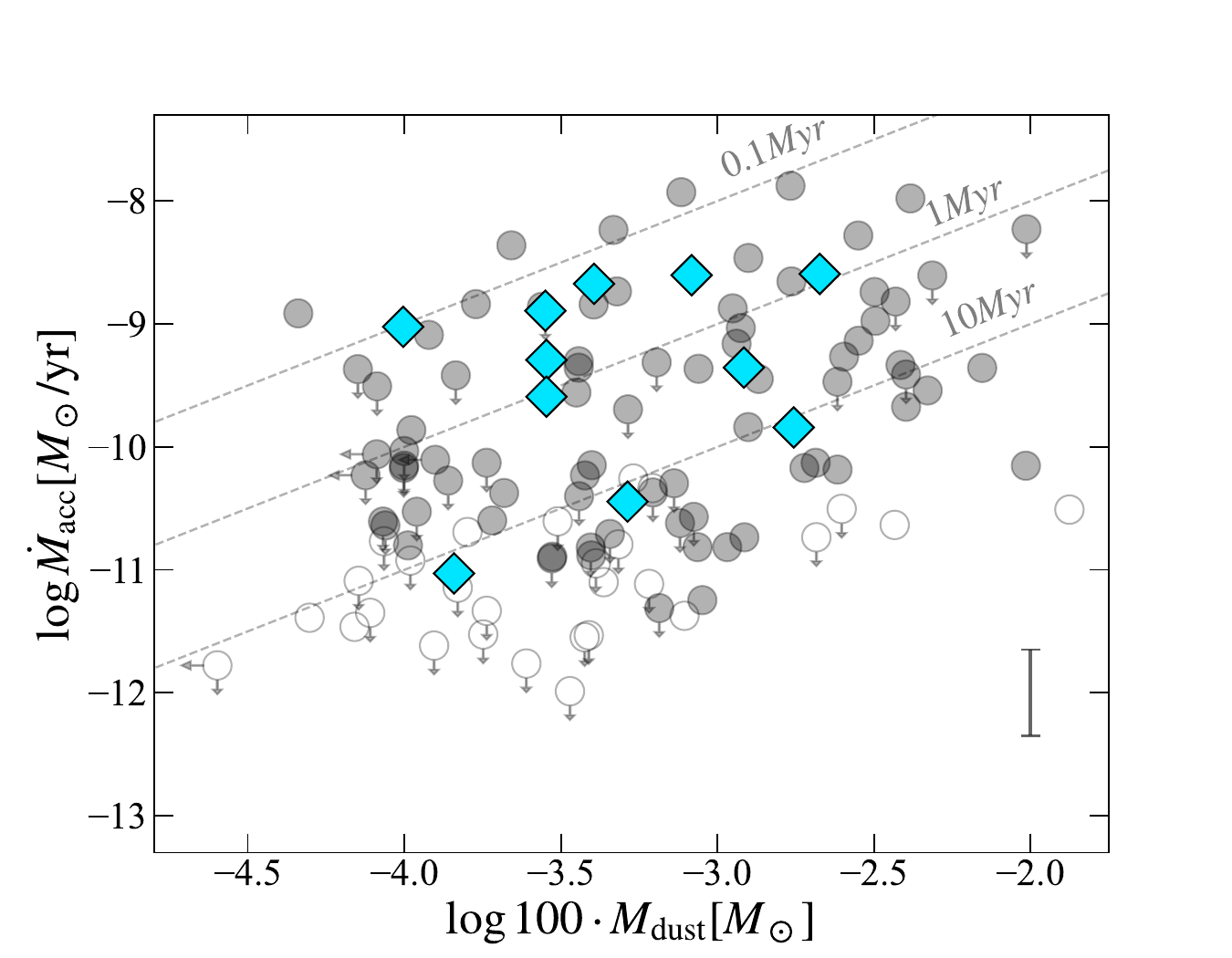}
  \caption{\macc\ as a function of \mdisc, with binary targets highlighted. Cyan diamonds represent known binaries present in the sample. We also indicate potential binary systems in the sample identified by \citet{Barenfeld2019} and Xie et al. (in prep). The grey error bar represents the typical uncertainty in \macc.}
     \label{fig: macc_mdisc_binaries}
\end{figure}

Previous studies have shown that different types of objects occupy different regions of the \macc-\mdisc\, plot and may partly explain the observed spread. For example, \citet{Zagaria2022} show that, using measured accretion rates from Lupus, Chameleon I and USco, discs in binary systems tend to have higher mass accretion rates, leading to observationally estimated $\tau_{\rm acc} \approx$ 0.1 Myr. In Figure \ref{fig: macc_mdisc_binaries}, we highlight objects in our sample that are either confirmed binaries (including those in \citealt{Zagaria2022}, mainly from \citealt{Barenfeld2019}) or suspected of having a close companion \citep[][Xie et al. in prep]{Barenfeld2019}. Although two points lie close to the 0.1 Myr accretion-timescale line, the sample here spans accretion rates spanning from $\sim 10^{-11} - 10^{-9} M_\odot / \rm{yr}$ and accretion timescales of $\tau_{\rm acc} \approx$ 0.1-10 Myr. The low sample size (11 objects) makes it difficult to draw firm conclusions, but the spread of these points suggests that the higher \macc\ values in the sample cannot be explained by available information on binarity. As the sample of binaries is certainly incomplete, future efforts to determine the binarity of the targets could better test this hypothesis.  

\begin{figure}
\centering
\includegraphics[width=0.49\textwidth]{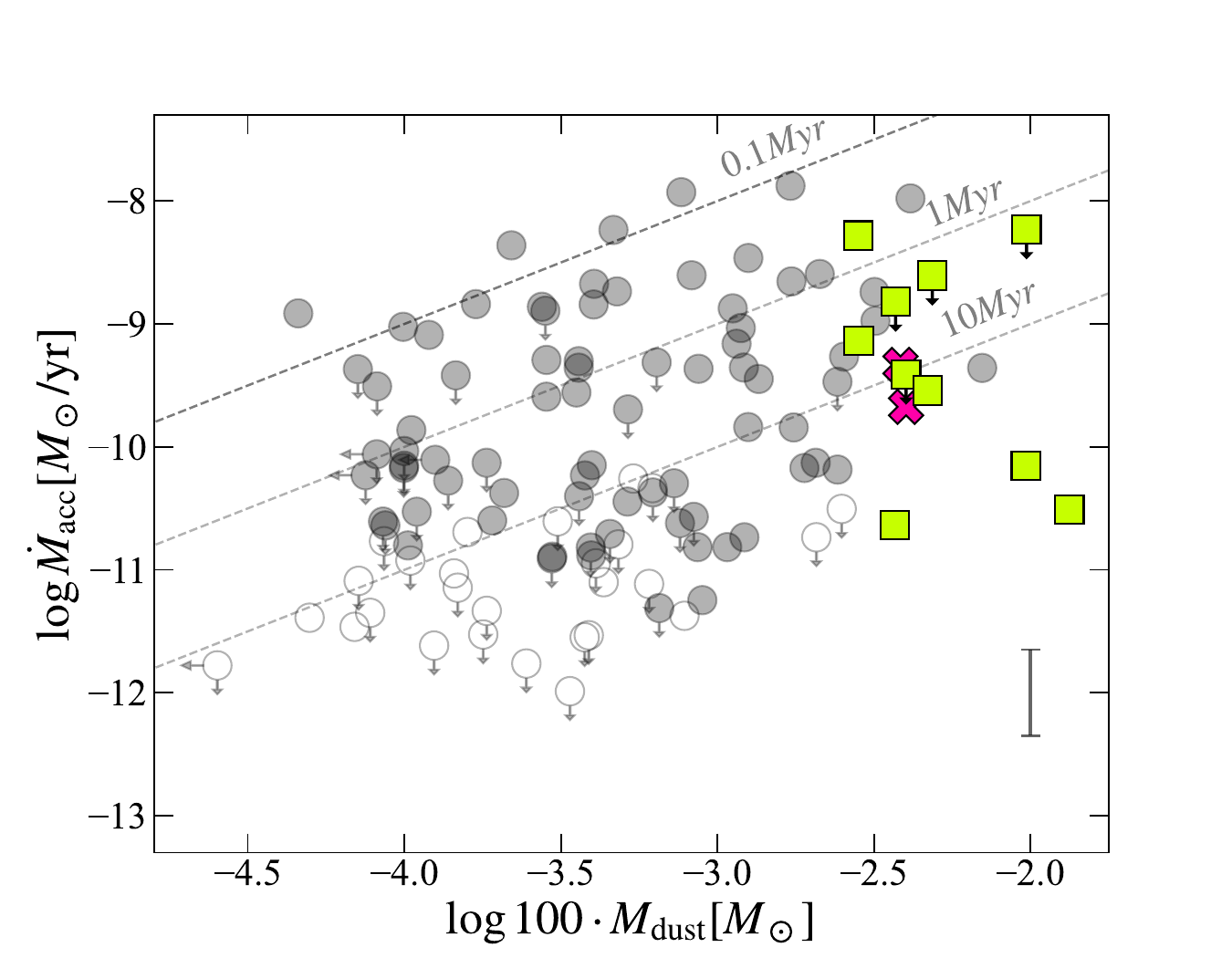}
  \caption{\macc\ as a function of \mdisc, with transition and structured discs. Transition discs are shown as lime squares and structured discs (evidence of disc substructures) as pink crosses \citep[see ][]{Carpenter2025, Pinilla2025}. The grey error bar represents the typical uncertainty in \macc.}
     \label{fig: macc_mdisc_tds}
\end{figure}

Transition disc objects are characterised by an optically thin inner disc (an inner hole or cavity, seen as a dip in the MIR of their spectral energy distribution) while maintaining excess emission at longer infrared wavelengths ($>10 \rm{\mu m}$) from the outer disc \citep[e.g.][]{Espaillat2014, vanderMarel2023}.  
\citet{Najita2007, Najita2015} demonstrated that transition discs occupy the higher \mdisc\ but lower \macc\ regions of such diagrams, as also observed in other studies \citep[e.g.][]{Manara2016b,Mulders2017}. On the other hand, the difference is smaller when considering the dependence of accretion on stellar mass \citep{Manara2014}. Follow-up studies comparing observations with planetary synthesis models show that the lower \macc\, at a given \mdisc\ could be attributed to the formation of giant planet(s) within the cavities of such discs \citep{Manara2019}. The identified transitional and structured discs in the sample \citep{Carpenter2025, Pinilla2025} have the highest disc masses (Fig.~\ref{fig: macc_mdisc_tds}). This is expected because of a selection bias, as we cannot yet resolve the inner cavities of less massive discs. The \macc\ values are again widely distributed within this sample. Interestingly, four of these objects have upper-limit measurements, suggesting that the true average \macc\ may be lower, although it would still not reach $\tau_{\rm acc} \sim $0.1 Myr. This suggests that the accretion properties of the transition discs do not differ from those of the other objects in the sample. Comparison with other star-forming regions (Figures \ref{fig: macc_mstar_pp7}, \ref{fig: macc_mdisc_pp7}) places these transition discs at intermediate \mdisc\ and \macc.

These results tentatively suggest that the extremes of the \macc\ distribution in USco cannot be explained by the general properties of binary and transitional-disc objects. However, both cases have small sample sizes, making it difficult to draw firm conclusions. Larger sample sizes with high-resolution ALMA data are required to better understand the effects of these objects in the \macc-\mdisc\ relation.

\subsection{Comparison of USco with other regions}\label{sec:comparison_others}

\begin{figure}
    \centering
    \includegraphics[width=0.49\textwidth]{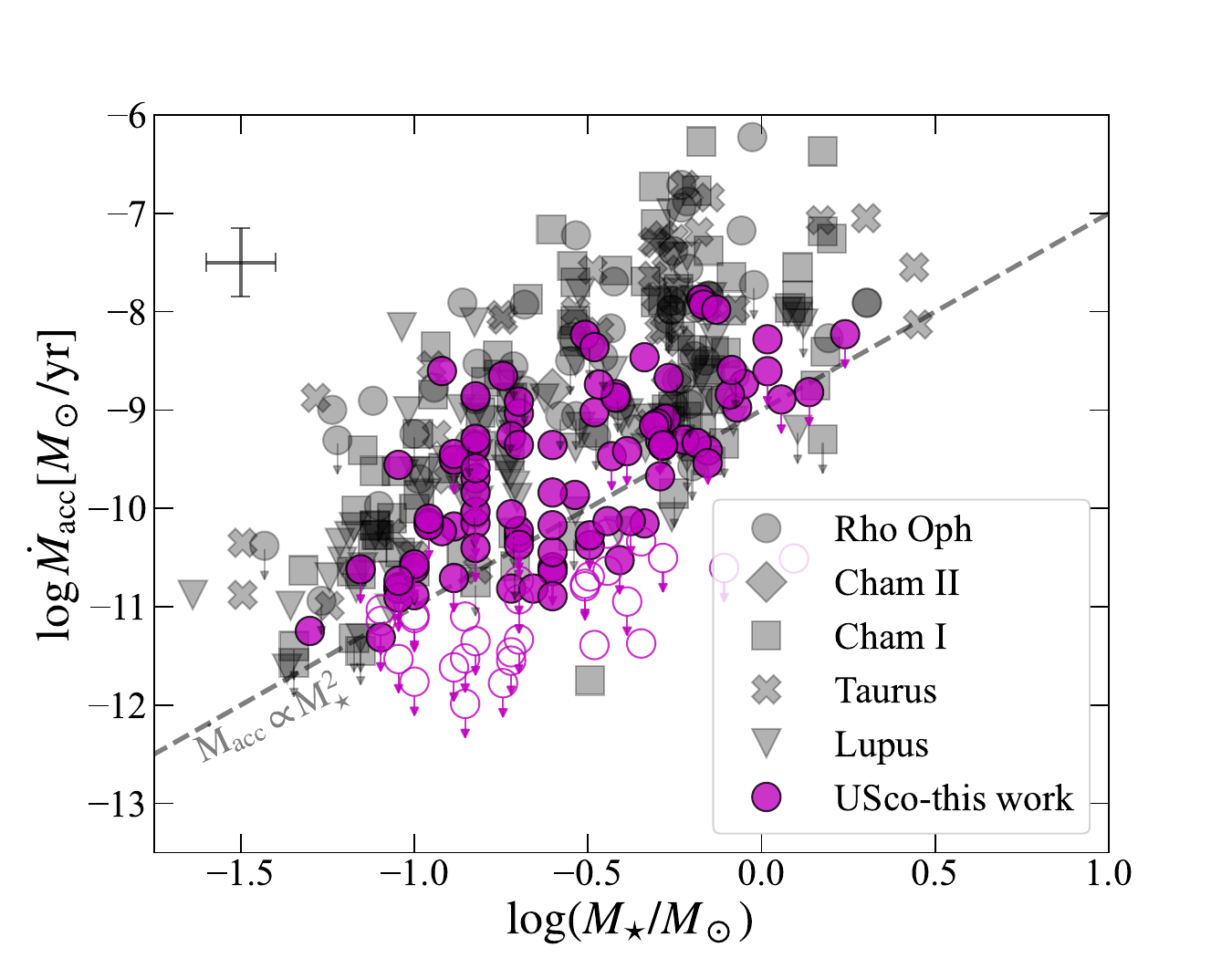}
    \caption{\macc\ as a function of \mstar\ for this sample, alongside values for the younger star-forming regions: $\rho$ Ophiuchus (L1688), Chamaeleon (Cham) I and II, Lupus, and Taurus. Data are from \citet{Manara2023}. The dashed grey line shows the relation $\dot{M}_{\rm acc} \propto M_\star ^2$ to guide the eye. The grey cross represents typical uncertainties in \macc,\mstar. }
    \label{fig: macc_mstar_pp7}
\end{figure}

Comparison of the best fit-power laws shown in Sect. \ref{sec: results} with those reported for other star-forming regions suggests that USco is similar to its younger counterparts. The dependence of \lacc\ on \lstar\ shows the strongest correlation in our sample, with $\rho = 0.5 \pm 0.1$ and $\beta = 2.1 \pm 0.4$. This dependence is similar to that reported for Lupus by \citet{Alcala2017}, who also report a lack of strong accretors at low stellar luminosities (\lacc $\leq$ 0.1 \lstar). The correlation between mass accretion rate and stellar mass in our sample is also similar to those measured in Lupus, $\rho$ Ophiuchus (L1688), and Chameleon I \citep[$\beta=$ 1.6, 1.8, and 2.3][]{Testi2022}. However, the lower $\alpha$ value of $-9.1 \pm 0.3$ in our sample suggests that the overall mass accretion rates are statistically lower, possibly reflecting the expected age dependence. This is illustrated in Fig. \ref{fig: macc_mstar_pp7}, where the values of this sample are plotted alongside those reported for the other regions by \citet{Manara2023}. 

\begin{figure}
    \centering
    \includegraphics[width=0.49\textwidth]{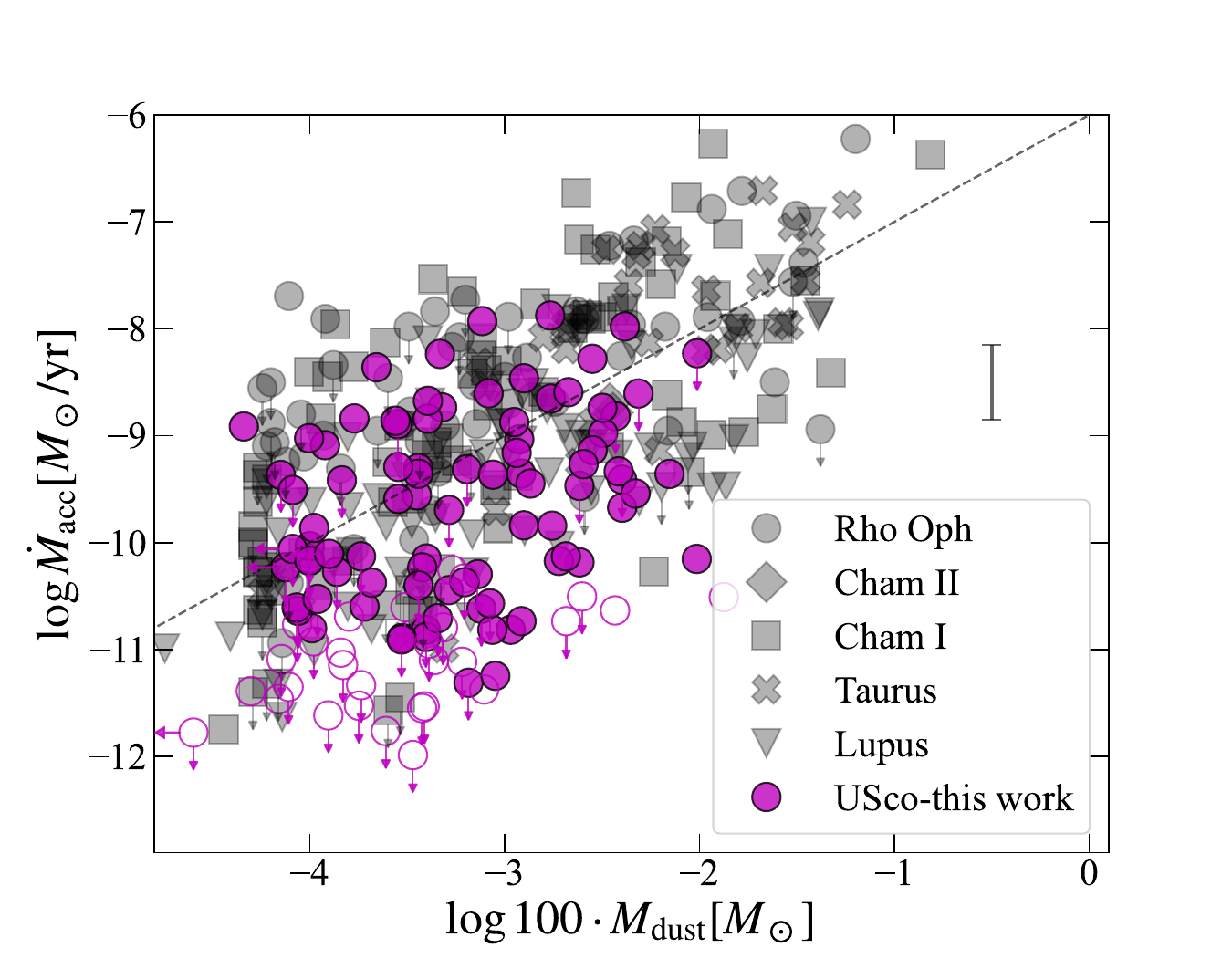}
    \caption{\macc\ as a function of \mdisc\ for this sample alongside values for the younger star-forming regions: $\rho$ Ophiuchus (L1688), Chamaeleon (Cham) I and II, Lupus, and Taurus. Data are from \citet{Manara2023}. The dashed grey line shows a disc lifetime ($\tau_{\rm acc} = M_{\rm disc} / \dot{M}_{\rm acc}$) of 1 Myr to guide the eye. The grey error bar represents the typical uncertainty in \macc.}
    \label{fig: macc_mdisc_pp7}
\end{figure}

\begin{figure}
\centering
\includegraphics[width=0.4\textwidth]{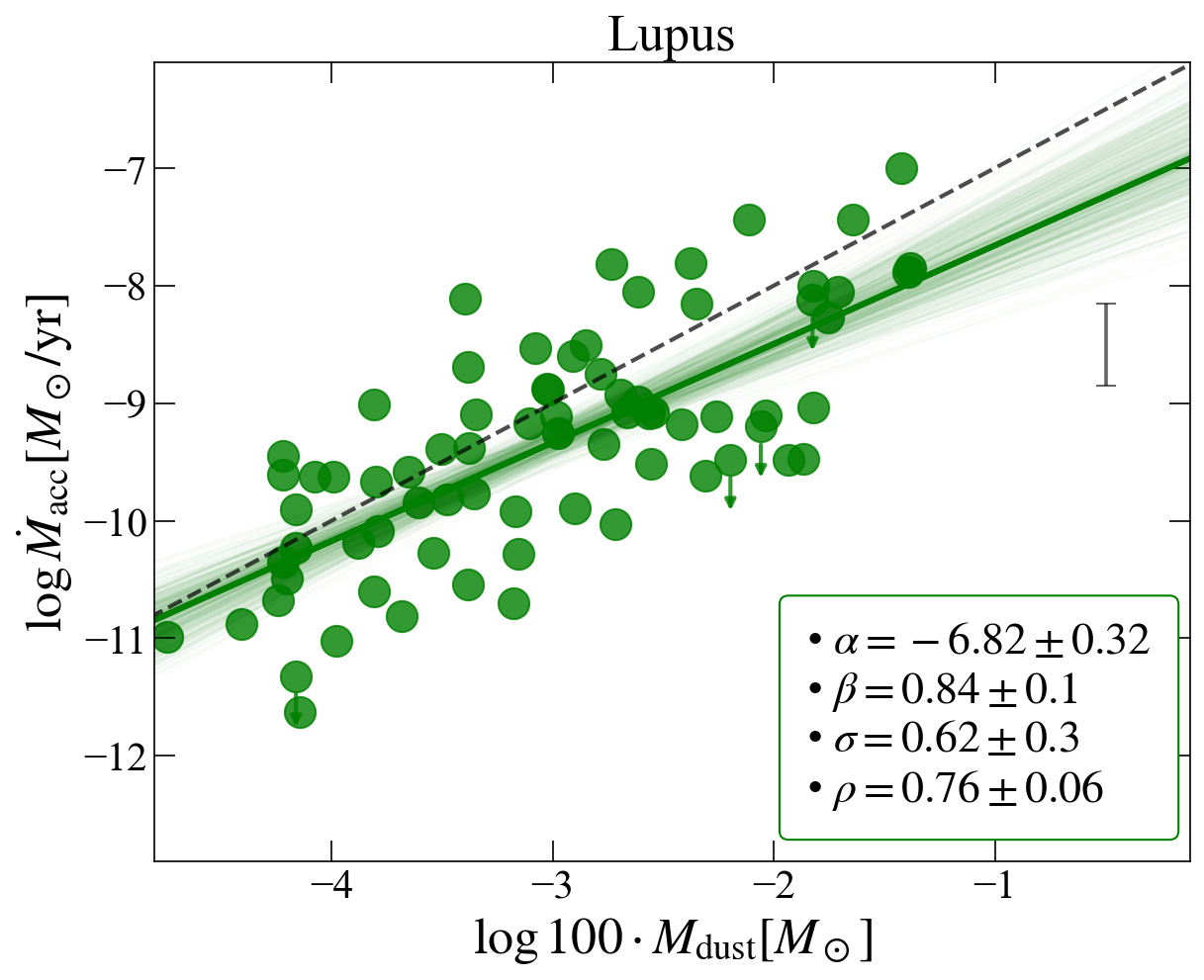}
\includegraphics[width=0.4\textwidth]{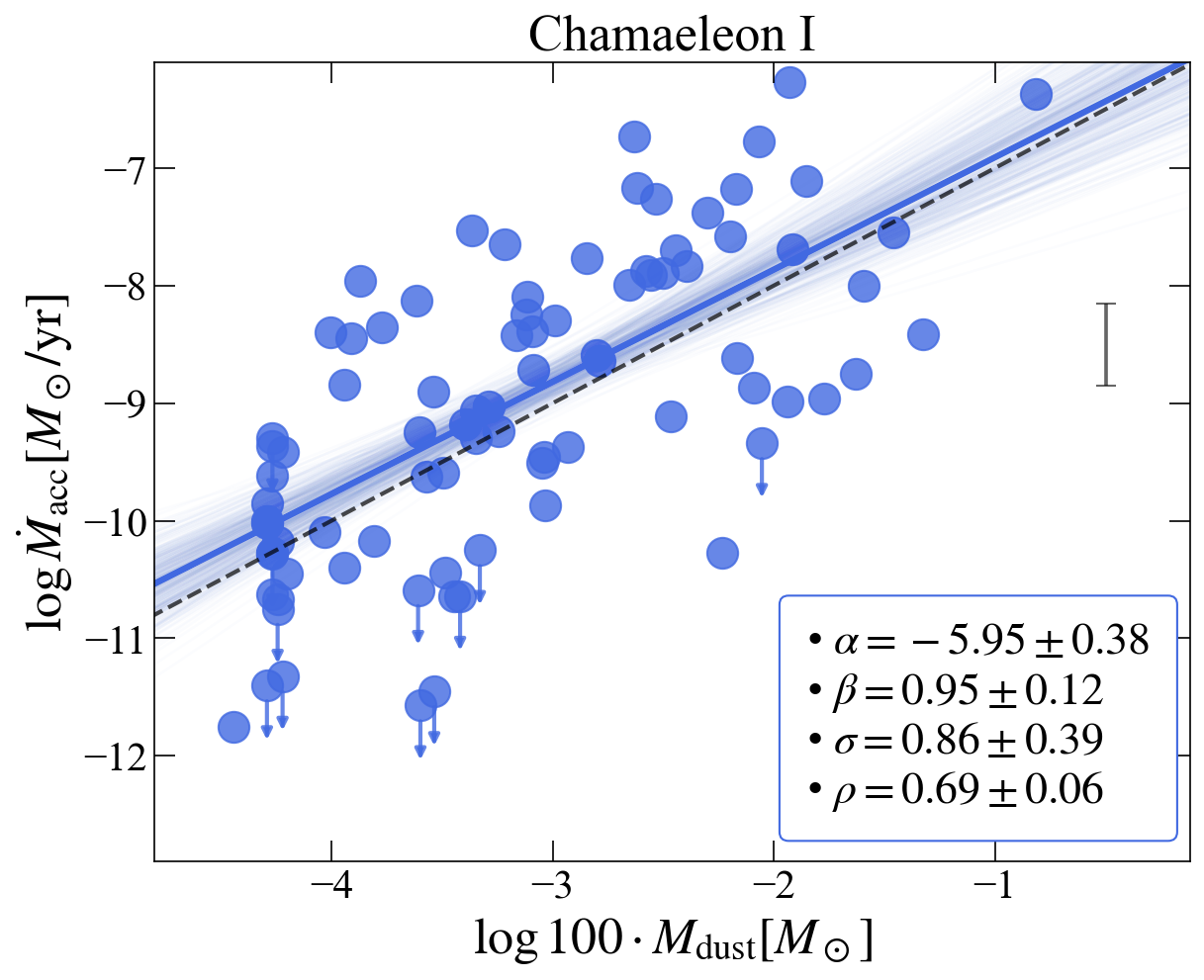}
\includegraphics[width=0.4\textwidth]{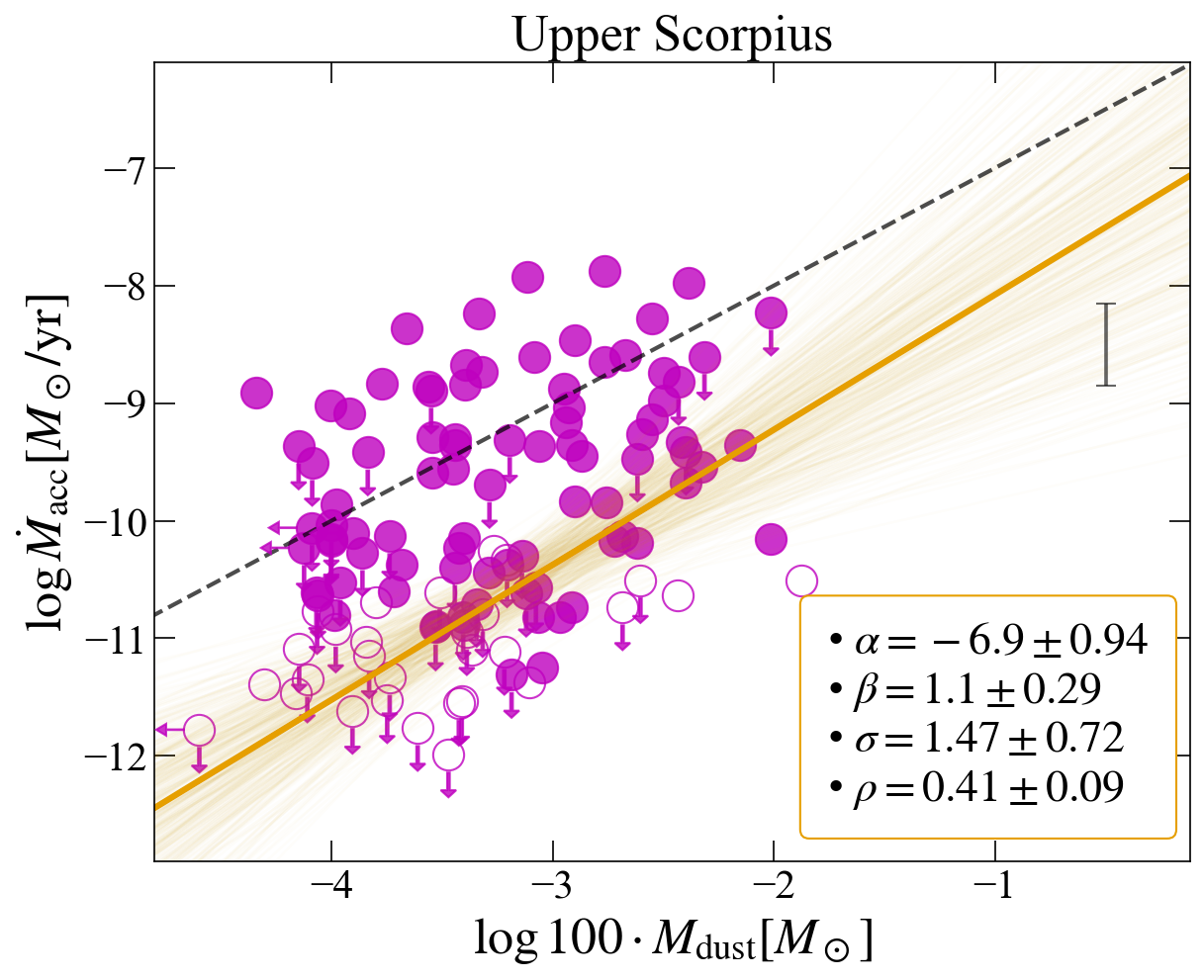}
  \caption{$\dot{M}_{\rm acc}$ as a function of $M_{\rm disc}$ for the Lupus, Chamaeleon I and USco star-forming regions. Values for Lupus and Chamaeleon I are taken from \cite{Manara2023} while the USco sample is presented in this work. In all panels the relations are fitted using the \textit{linmix} python package \citep{Kelly2007}. The fit parameters for each region are shown in the text boxes of each subplot. The dashed black line represents the ratio $\tau_{\rm acc} = 1.0$ Myr. The grey error bar represents the typical uncertainty in \macc.}
     \label{fig: macc_mdisc_regions}
\end{figure}

Some similarities extend to the \macc-\mdisc\ relation. Figure \ref{fig: macc_mdisc_pp7} shows that the measurements are generally consistent with those of the other regions and show no obvious separation with age. However, the strongest accretors in our sample have lower \macc values than those in other regions. The range of measured disc masses is also smaller in USco, with a cut-off at $\sim 10^{-2} M_\odot$. Figure \ref{fig: macc_mdisc_regions} illustrates the similarity between the best-fit slopes in Lupus, Chameleon I and our USco sample; all of which have $\beta \sim 1.0$. However, although the correlation coefficient remains similar in the two younger regions ($\rho \sim 0.7$), it is significantly weaker in USco ($\rho \sim 0.4$), indicating a significant shift in the strength of the correlation.

Although the best-fit \macc\ on \mstar\ and \mdisc\ relations in USco are not indifferent to those found in the younger star-forming regions, the large spread in the data suggests a different story. A consistent feature of the results is the large dispersions in the measured properties of USco. Figures \ref{fig: macc_mstar} and \ref{fig: macc_mdisc} show, for example, that in a given \mstar\ or \mdisc\ bin, \macc\ varies by more than four orders of magnitude. Similar large spreads are observed in the \lacc\ - \lstar\ and \macc\ - \Rco\ relations. The significance of these are reflected in the poorly measured correlations, all of which have $\rho \le 0.5$. Moreover, the \macc\ on \Rco\ relation is the weakest in the sample. A $\beta$ value of $0.0 \pm 0.8$ together with $\rho = 0.0 \pm 0.1$ indicates that these two measurements are not related. Across all relations, the magnitude of the spreads in USco are at least equal to, and generally larger than, those reported for younger regions \citep{Alcala2014, Alcala2017,Testi2022}, whereas the correlations are weaker (Fig.~ \ref{fig: macc_mdisc_regions}). Together with the larger dispersion ($\sigma$), this indicates that we are witnessing the disappearance of the relations in USco, particularly between \macc\ and \mdisc. The lack of a clear dependence of \macc\ on \mdisc\ or \Rco\ suggests that the amount of material in the disc has no relevance to the amount of ongoing accretion implying a true disconnect between the inner and outer discs at the USco stage. 

\subsection{Implications for disc evolution}

The accretion and disc properties in USco are believed to be an important factor in disentangling which mechanisms govern the evolution of protoplanetary discs. Although an in-depth analysis of the implications for evolutionary models is beyond the scope of this paper, we discuss briefly discuss our results in this context. 

A key feature of the results in USco is the large spread of accretion rates and the weak correlations between accretion and the stellar or disc properties. This result, particularly when shown to be larger than that observed in the younger star-forming regions, contrasts with the expectations of viscous evolution theories. Under this prescription, the spread in the relations is expected to diminish over time, with values converging onto their theoretical isochrones \citep{Lodato2017,Somigliana2024}. This also predicts a changing slope in the power laws with the age of the region, which is not what we observe when we compare our sample with the younger regions. It further suggests that viscous theory alone cannot explain what we see in USco \citep[e.g.][]{Anania2025}. 

Alternatively, wind-driven models reproduce the large spreads observed in USco more successfully. Following the initial tests by \citet{Mulders2017} to reproduce the observations in Lupus, \citet{Tabone2022} showed that MHD wind-driven models reproduce the observed \macc\ - \mdisc\ spread in Lupus. This was later also tested in USco, but the agreement was dependent on the initial conditions. \citet{Tabone2025} find that reproducing the trends observed in the AGE-PRO survey requires initially compact discs with medium mass-loss rates and short accretion timescales. In contrast to viscous models, \citet{Somigliana2024} show that the slopes of the power-law relations remain the same, but that the shapes change in the wind-driven scenario. These models also reveal a larger spread in the disc lifetimes at later stages. Although neither approach reproduces the general properties of all the observed star-forming regions, the wind-driven model provides a better match to the M20 results and is more consistent with our observations. Furthermore, although wind-driven models can somewhat reproduce the lack of a correlation in the relations, they currently do not explain how one can diminish over time.

Additional processes, such as external photoevaporation \citep{Rosotti2017}, must also be considered to better account for the measured properties in the region. The moderate levels of far-ultraviolet (FUV) flux experienced by USco discs \citep[approximately 2-12 $G_0$; ][]{Anania2025} have been proposed to explain the smaller average gas disc radii (Zagaria et al. in prep). Hybrid models that include the effects of photoevaporation reproduce the disc gas sizes and disc fractions in USco more successfully.  However, they still struggle to simultaneously predict accretion rates and disc masses. The evolution of dust properties is another important factor, as suggested by M20. The results from the AGE-PRO large programme show that gas and dust properties evolve differently, and that dust mass is not an ideal tracer of disc mass, particularly at older ages. Incorporating dust evolution into the viscous models of \citet{Sellek2020} improved agreement with the observed spreads in mass accretion rate and disc dust mass in USco and Lupus. Nevertheless, the current models fail to explain the full range of properties measured in USco, particularly the large spreads that we report here. Further work is required to include photoevaporation and dust evolution in evolutionary models.

Overall, the results for USco challenge the predictions of both viscous and wind-driven evolutionary scenarios. In particular, the large spreads observed in the region, characterised by its older age, is a standout feature that requires further modelling efforts to understand. In addition, a more complete sample of the gas properties of USco discs is needed to unravel the different evolutionary scenarios.
   
\section{Summary and conclusions}\label{sec: conclusions}

In this work we present the results of a spectroscopic survey measuring the stellar and accretion properties for 127 disc-bearing stars in the USco star-forming region. By analysing their X-Shooter spectra with the multicomponent fitting code FRAPPE \citep{Claes2024}, we derive the stellar properties together with their accretion luminosity and mass accretion rates from the UV continuum excess emission. By combining these measurements with ALMA flux densities and CO disc radii \citep[][Zagaria et al. in prep]{Carpenter2025}, we present the first large catalogue of USco stellar, accretion, and disc properties. We performed a statistical analysis of several key relations between these properties and fitted them with power laws, placing the results in the context of those from younger star-forming regions. We summarise the main findings of our work below:

\begin{itemize}
    \item We applied a novel approach to determine upper limits on the accretion luminosities based on the S/N of the Balmer continua in the X-Shooter spectra. By calculating +1$\sigma$ upper-limit value on \lacc\, we systematically distinguish between definite accretors and those with upper limit measurements. 
    \item Approximately 50\% of the objects have a clearly measured accretion luminosity from the UV excess emission (definite accretors). For the remaining 50\% we can only determine an upper-limit on \lacc (upper-limit accretors). This is due to the excess being intrinsically very low, non-existent, or because the S/N of the observations is poor. For a number of objects, the measured excess is below the value expected from chromospheric emission for stars of the same spectral type. These upper-limit accretors are distributed across the full range of \mstar, \lstar, and \mdisc\ values.  
    \item We analysed the relations between accretion and stellar properties. A key result is the large spread (approximately four orders of magnitude in \lacc), with a number of strong accretors ($L_{\rm acc} \geq 0.1 \cdot L_\star$) along with low upper-limit accretors. Power-law fits yield low values of the correlation coefficient ($\rho \sim 0.4-0.5$) in the \lacc-\lstar\ and \macc-\mstar relations. In addition, analysis of the dependence of accretion on disc properties (dust mass and gaseous disc radii) shows that these correlations are even weaker.
    \item The relations between accretion and stellar properties are, in general, consistent with those of other star-forming regions (e.g. Lupus, $\rho$ Ophiuchus, and Chamaeleon I). However, the spreads observed in USco are significantly larger than in its younger counterparts, especially when the disc properties are included. Together with the absence of a correlation between \macc\ and disc dust mass or radius, this suggests a fading of the relations with age and that the inner and outer disc become increasingly decoupled by the age of USco ($\sim$5-10 Myr).
    \item These results could not be explained by membership of the objects to the various USco subgroups \citep{Ratzenbock2023}. Similarly, the presence of (known) transition discs and binary systems was found to not affect the results. 
\end{itemize}

The wealth of information contained in the X-Shooter spectra of this sample will provide more information on the properties of the USco region. Ongoing work (Empey et al. in preparation) will characterise the permitted and forbidden emission lines to further shed light on the accretion properties, chromospheric contamination, and the strength and origin of disc winds. Future work should focus on characterising the gas properties of disc-bearing stars in USco to further build upon the current understanding of the region and its role in the overall evolutionary pathway. Complementing this with additional modelling of dust evolution and photoevaporation alongside evolutionary models, would also be highly beneficial.

\section*{Data Availability}
Tables \ref{tab: master_props} and \ref{tab: logobs} are only available in electronic form at the CDS via anonymous FTP at \url{cdsarc.u-strasbg.fr} (130.79.128.5) or via \url{http://cdsweb.u-strasbg.fr/cgi-bin/qcat?J/A+A/}. 

\begin{acknowledgements}
We thank the anonymous referee for their careful consideration and comments that helped improve the clarity of our work. A.E. acknowledges funding from Taighde Éireann – Research Ireland under Grant number GOIPG/2023/4396 and the UCD Physics Scholarship in Research and Teaching (SIRAT). A.E. acknowledges support from the ESO Early-Career Scientific Visitor Programme when part of this work was carried out. CFM and KM are funded by the European Union (ERC, WANDA, 101039452). Views and opinions expressed are however those of the author(s) only and do not necessarily reflect those of the European Union or the European Research Council Executive Agency. Neither the European Union nor the granting authority can be held responsible for them. NOIRLab IRAF is distributed by the Community Science and Data Center at NSF NOIRLab, which is managed by the Association of Universities for Research in Astronomy (AURA) under a cooperative agreement with the U.S. National Science Foundation. GL acknowledges support from PRIN-MUR 20228JPA3A, from the European Union Next Generation EU, CUP:G53D23000870006. JMA and BN acknowledge financial support from Large Grant INAF-2024 “Spectral Key features of Young stellar objects: Wind-Accretion LinKs Explored in the infraRed (SKYWALKER).

\end{acknowledgements}

\bibliographystyle{aa}  
\bibliography{ref}

\clearpage
\begin{appendix}

\section{Derivation of upper limits}\label{sec: uplims_details}
Among the USco sample we record a significant number of very low values of accretion luminosity measured from their UV continuum excess with FRAPPE. In many cases these are on or below the sensitivity limit of what can be reliably measured from the X-Shooter spectra. To allow for an accurate statistical analysis of the region, there is a need to distinguish between definite measurements of excess due to accretion and those which are upper limits. In this study we develop a novel approach to separating such measurements based on the S/N ratio in the Balmer continua of the observed spectra, while also calculating +1$\sigma$ upper limits on the accretion luminosity. In this section we describe the procedure we follow to do this. \\

For each object we first correct the X-Shooter spectrum for interstellar extinction using the best fit value of $A_V$ determined by FRAPPE, and the Cardelli extinction law \citep{Cardelli89} with $R_V = 3.1$. With this we compute the mean value and a 1 $\sigma$ RMS for four bins in the Balmer continuum ($\lambda\lambda$ 332.5 - 337.5 nm, 337.5 - 342.5 nm, 352 - 358 nm, 359 - 362 nm). These are the same wavelength ranges used by FRAPPE in the fit to constrain the accretion excess. A relative uncertainty was then defined by taking the average ratio of the +1$\sigma$ RMS to the mean flux across all bins. The uncertainties associated with the interpolated class III template fitted by FRAPPE are negligible compared to the noise in the observed spectra, therefore they not considered in this analysis. This final value represents the uncertainty on the flux of the observed Balmer continuum, and therefore that of the excess due to accretion. This fractional excess was applied to the observed (uncorrected) Balmer continuum flux (332.5-362.0 nm) to create a +1$\sigma$ artificial spectrum. For each object the analysis with FRAPPE was reran on the artificial spectrum with a fixed input SpT and $A_V$ (those initially derived, and reported in Table \ref{tab: master_props}). This leaves the excess due to accretion as the only free parameter in the fitter, and hence from the resulting fit we obtain a +1$\sigma$ value of the accretion luminosity. For each target we compare this with the original value of \lacc. A given object was determined to be an upper-limit measurement if the relative difference between the measured accretion luminosity ($L_{\rm acc}^{\rm M}$) and the one obtained artificially (+1$\sigma$), as just described, ($L_{\rm acc}^{\rm UL}$) exceeded 20\%. Mathematically this is: 

\begin{equation}\label{eq: uplims}
    \frac{L_{\rm acc}^{\rm UL} - L_{\rm acc}^{\rm M}}{ L_{\rm acc}^{\rm M}} > 0.2 .
\end{equation}

The choice of 20\% as a threshold was determined empirically by assessing a range of values and comparing the resulting classifications (upper-limit or definite accretion measurements) with a visual inspection of the FRAPPE fits. A value of 20\% provided the best agreement, minimizing both false upper limits and misclassified definite accretors. As a secondary check, we also inspect the equivalent widths (EW) of the $H_\alpha$ and compare them to the classical threshold of 10 \AA. Of the targets showing a clear excess, 86\% (61/71) have EWs $\geq$10\AA; nine of the remaining ten have EW $< 10$\AA ~while one target doesn't have a clear emission line profile (2MASS J116101264-2104446). For upper-limit accretion measurements, just under half (45\%, 22/49) have EWs $\geq 10$\AA, with all the others showing EWs < 10 \AA. Objects which were deemed to have upper-limit measurements by the method described above, are indicated by a 'y' in the upper limit (UL) column of Tab. \ref{tab: master_props}, and an 'n' if they show a reliable continuum excess. \\

\begin{figure}
\centering
\includegraphics[width=0.4\textwidth]{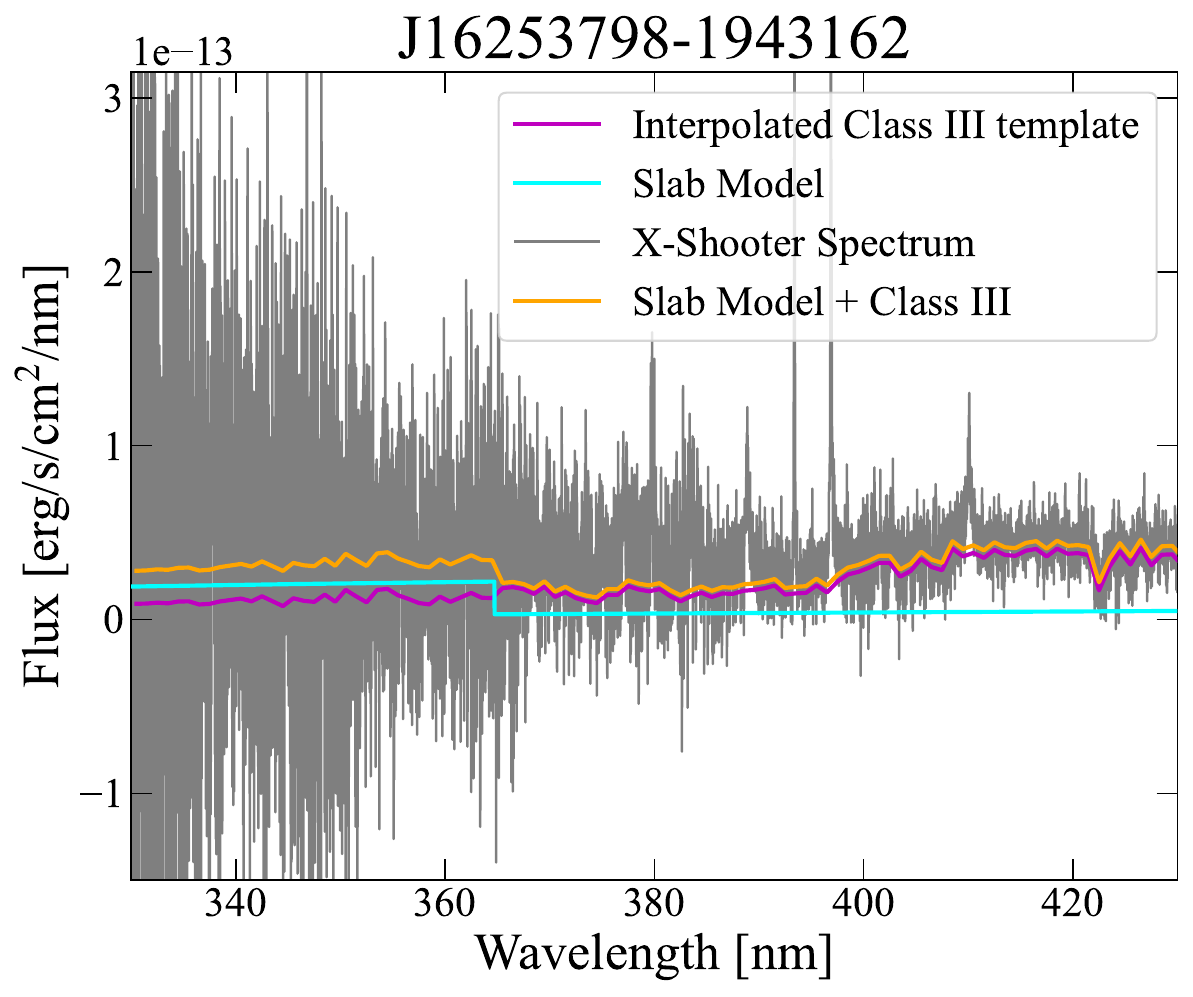}
\includegraphics[width=0.4\textwidth]{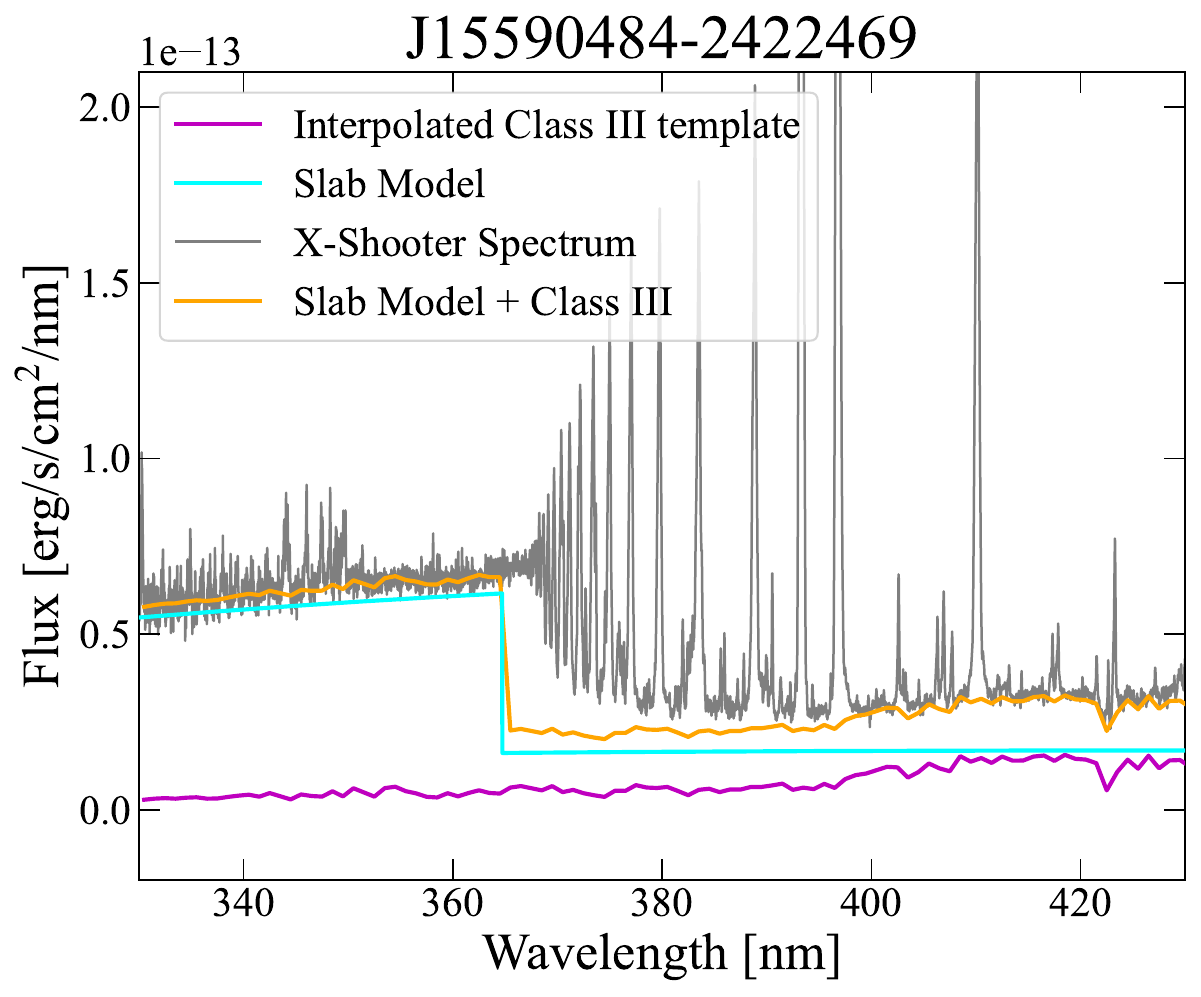}
\caption{Examples of the FRAPPE fits for upper-limit and definite accretors. In both plots, the de-reddened X-Shooter spectrum is plotted in grey, the interpolated class III template in magenta, in cyan the best fit isothermal hydrogen slab model, and in orange the class III template + slab model. On the left is an object defined as an upper-limit accretor based on its low S/N in the UV. On the right is a definite accretor where the UV continuum excess is clear. }
\label{fig: upperlimit_examples}
\end{figure}

Under the approach described above, a measurement of \lacc ~ has an upper-limit detection if either its spectrum has a poor S/N in the UV continuum, leading to difficulty in detecting accretion excesses or there remains no clear excess even with a good S/N. These are important factors to consider when evaluating the statistics of the sample (see Sect. \ref{sec: results}). Following this method we find that 49/121 targets have upper-limit accretion measurements (these are indicated by a "y" in the upper limit (UL) column of Table \ref{tab: master_props} and identified with a downward pointing arrow in any plots of \lacc\ or \macc. Figure \ref{fig: upperlimit_examples} shows the X-Shooter spectrum and FRAPPE fits for two examples, one for a upper-limit measurement (J16253798-1943162) and one for a true detection (J15590484-2422469). 

\section{Stellar variability} \label{sec: vars}
As described in Section \ref{sec: data_reduction}, the synthetic magnitudes of all the objects in the sample were compared with historical photometric measurements. Most were in good agreement; however, a sub-sample of seven targets showed significant differences. These differences are believed to be due to variability of the central star. All seven show g or V band variability of at least 1 magnitude in their lightcurves on the ASAS-SN Sky Patrol database \citep{Shappee2014, Hart2023}. Two objects (2MASS J16222160-2217307, 2MASS J16253849-2613540) were additionally classified as dippers and one as a stochastic variable star (2MASS J16020429-2231468) by \citet{Cody2018}. \\

We exclude just the latter object from the analysis due to uncertainty in its derived properties. While the comparison with photometry shows a difference of around 0.6 mag in the Gaia filters, the results from FRAPPE are also uncertain. We measure an extremely low \lstar\, for its associated \teff, which does not fall within the available evolutionary tracks on the HR diagram (Fig. \ref{fig: hr_diagram}). \citet{Cody2018} measured a particularly high variability amplitude (0.91 of its normalised flux). The ASAS-SN light curve showed maximum variations of 3 magnitudes in the g filter. It is possible that this object was caught during an active period while undergoing some form of unpredictable variability. Additionally, having a highly inclined disc can result in an obscured photosphere, leading to an underestimated value of its stellar luminosity \citep{Alcala2014}. Follow-up observations would be required to verify if this is the case for this particular target. To avoid these uncertainties influencing other results, it was discounted in the analysis of the accretion and stellar properties.

\section{Additional plots}\label{sec: add_plots}

\begin{figure*}
\centering
\includegraphics[width=0.49\textwidth]{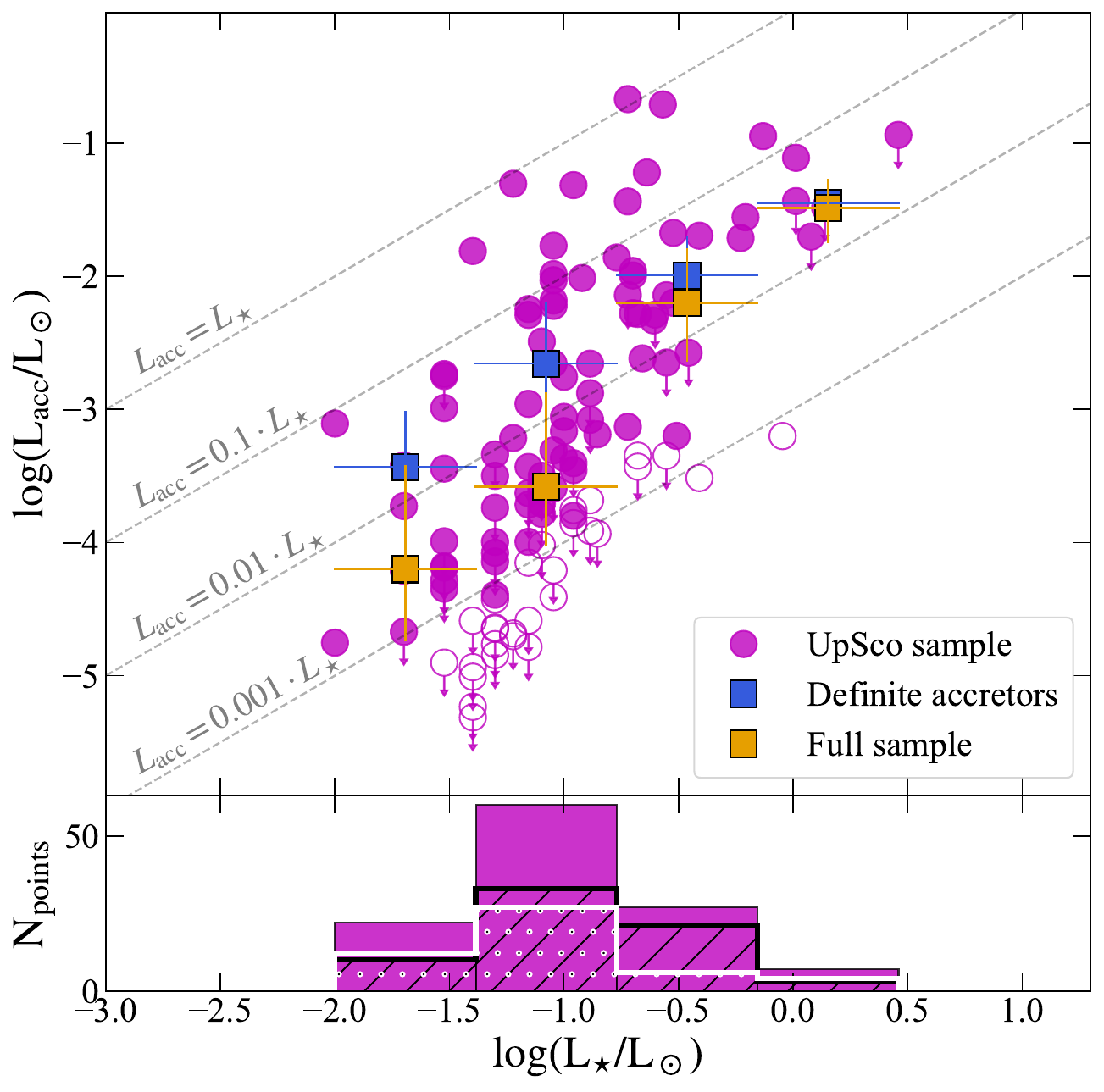}
\hfill
\includegraphics[width=0.49\textwidth]{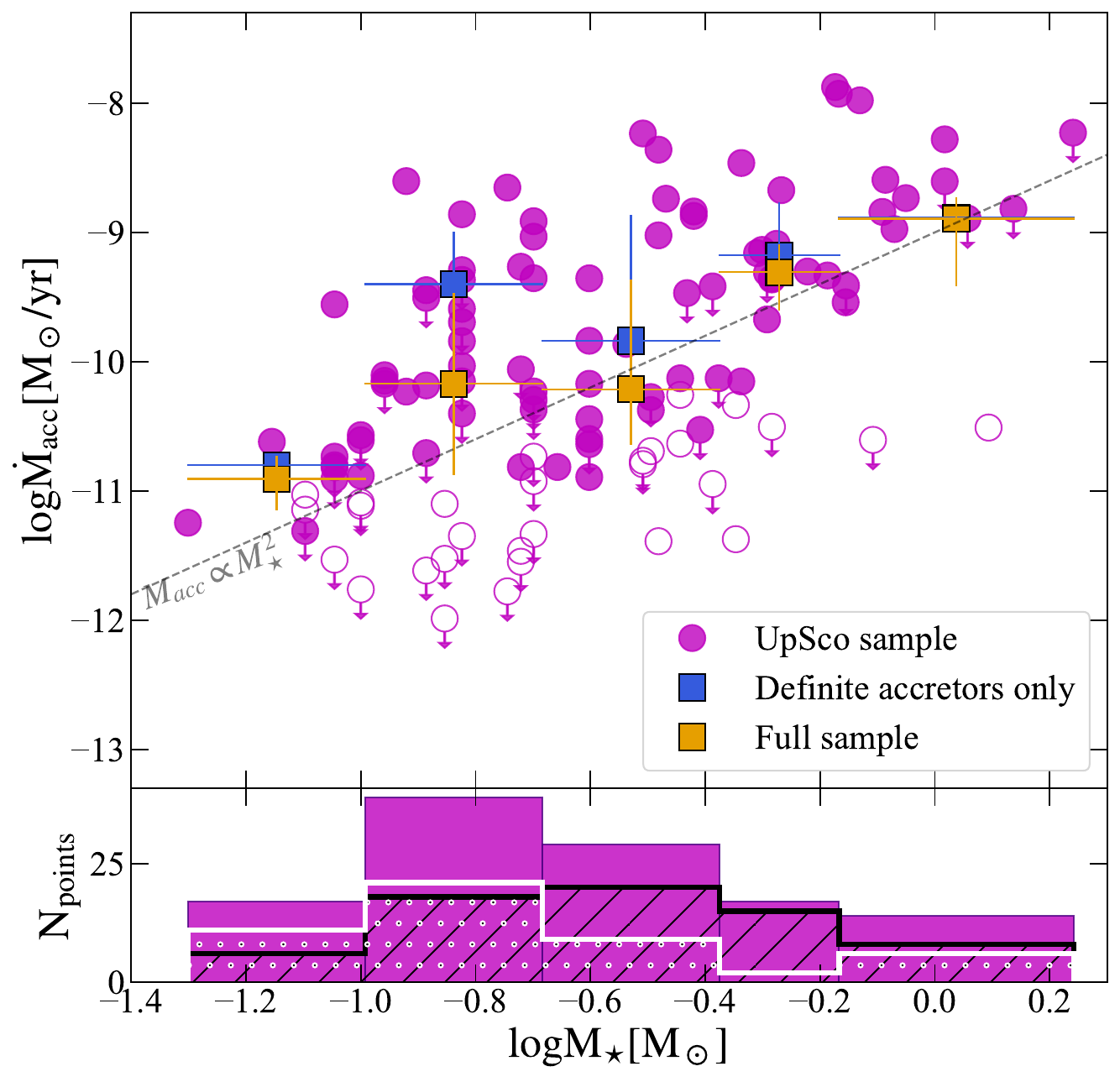}
\vspace{0.05 cm}
\includegraphics[width=0.49\textwidth]{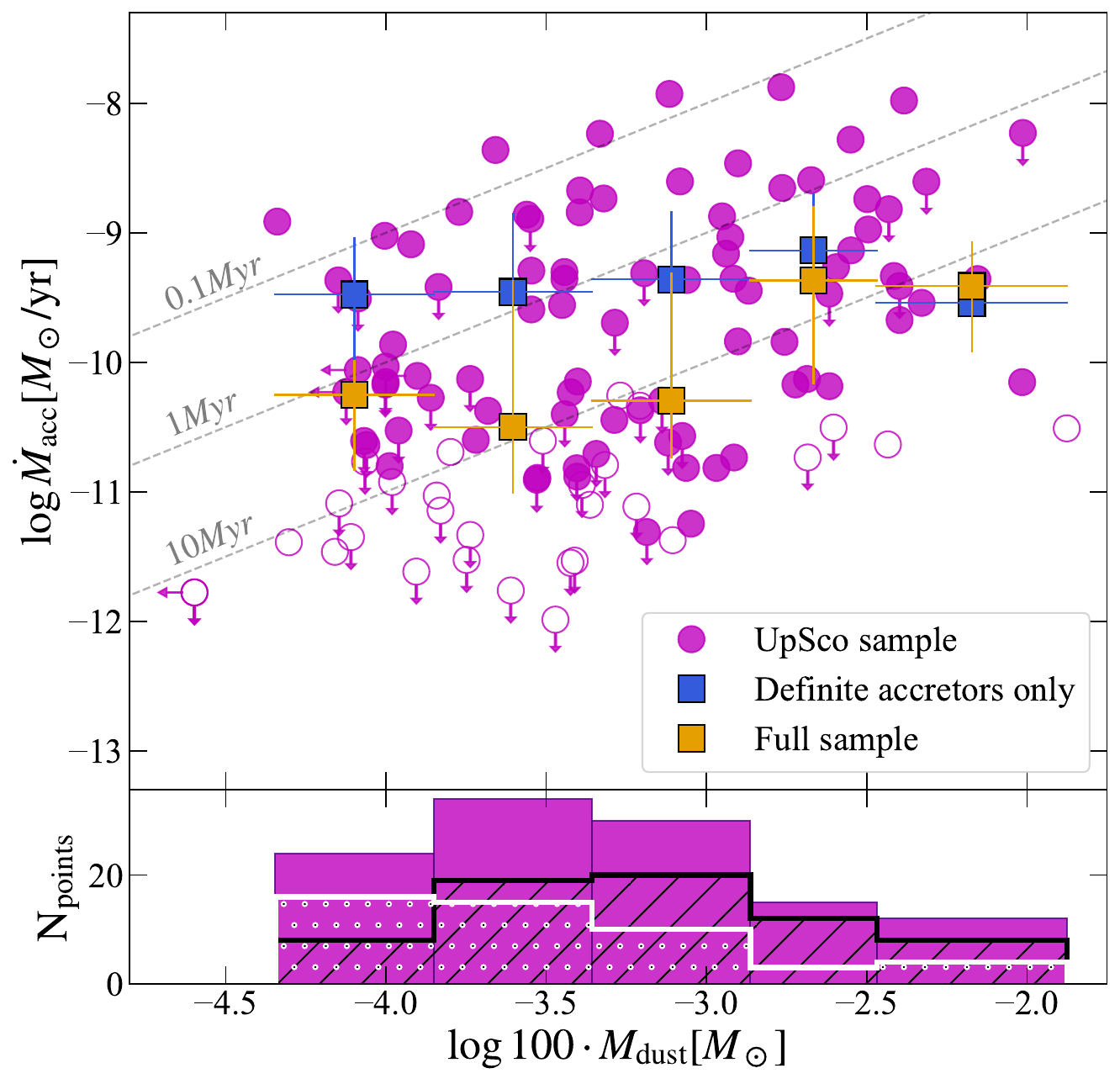}
\hfill
\includegraphics[width=0.49\textwidth]{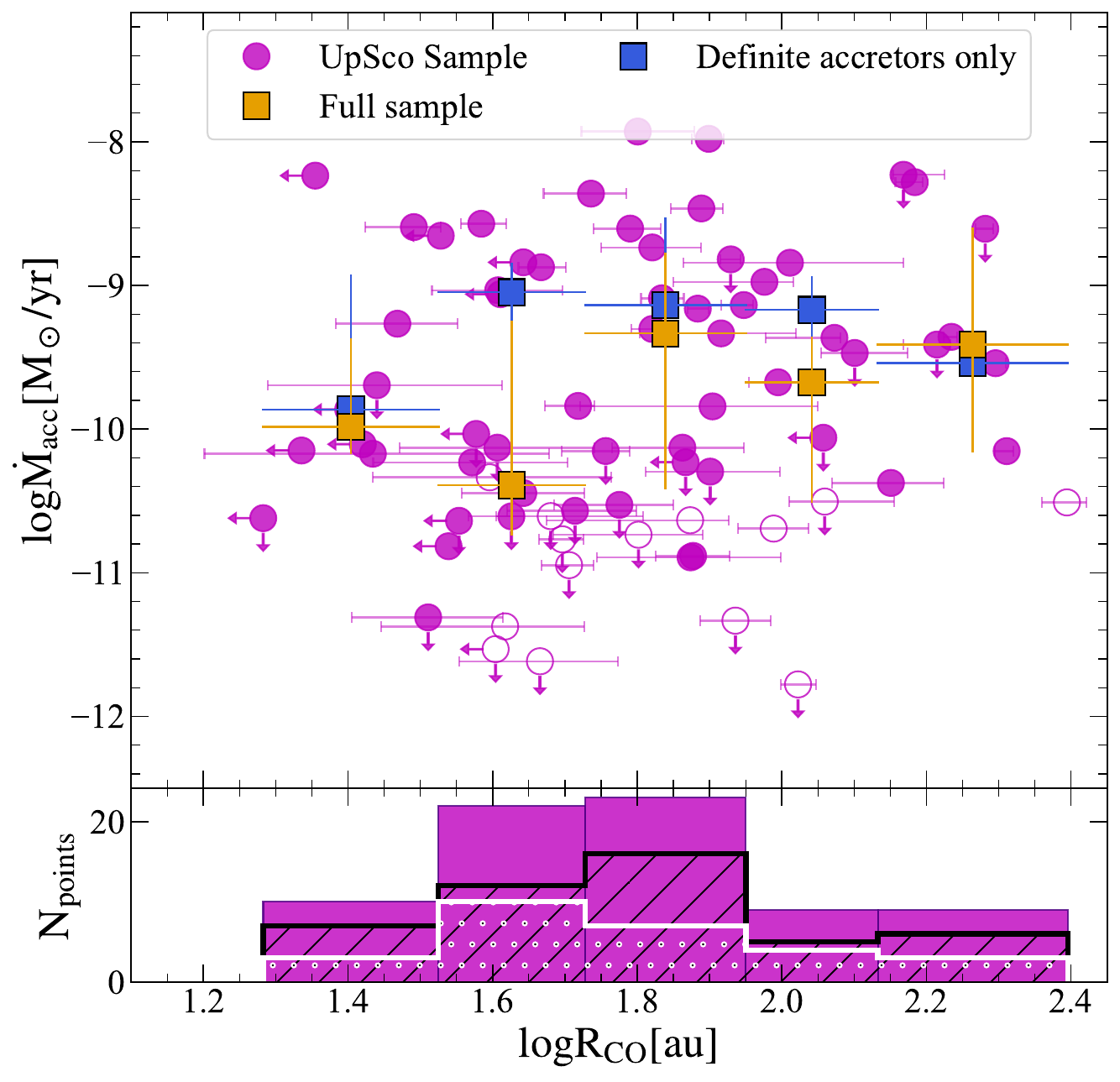}
\vspace{0.05 cm}
\caption{Scatter plots and histograms of stellar, disc and accretion properties for the USco sample. Squares represent the medians while vertical lines represent the limits of the 25th and 75th quantile in a given bin of \lstar, \mstar, \macc, \mdisc, and \Rco. Horizontal lines are indicative of the width of given bin. Blue medians and quantiles are computed for definite accretors only (solid points, no arrow), orange squares and lines are for the full sample (including upper limits, treated equally). We note for bins where the number of upper-limit accretors exceed the number of definite accretors the median and quantile values are not well defined. Bottom panel: histograms of the full sample (magenta), upper limits only (white dotted), and detections only (black striped) in each bin. . From top left: \lacc vs \lstar, \macc vs \mstar, \macc vs \mdisc, and \macc vs \Rco.}
\label{fig: binned_hists}
\end{figure*}

For each relation considered in Section \ref{sec: results}, as well as considering the scatter we also compute medians, 25th and 75th quantile values for bins of the independent variable. This is done while considering define accretors only and also for the whole sample (including the upper-limit accretors equally). We note that where the number of upper-limit accretors in a given bin exceeds those of the definite accretors, the true median and lower quantile are not well defined. The results of this approach are  seen in Figure \ref{fig: binned_hists}.

\begin{figure*} 
    \centering
    \includegraphics[width=0.49\linewidth]{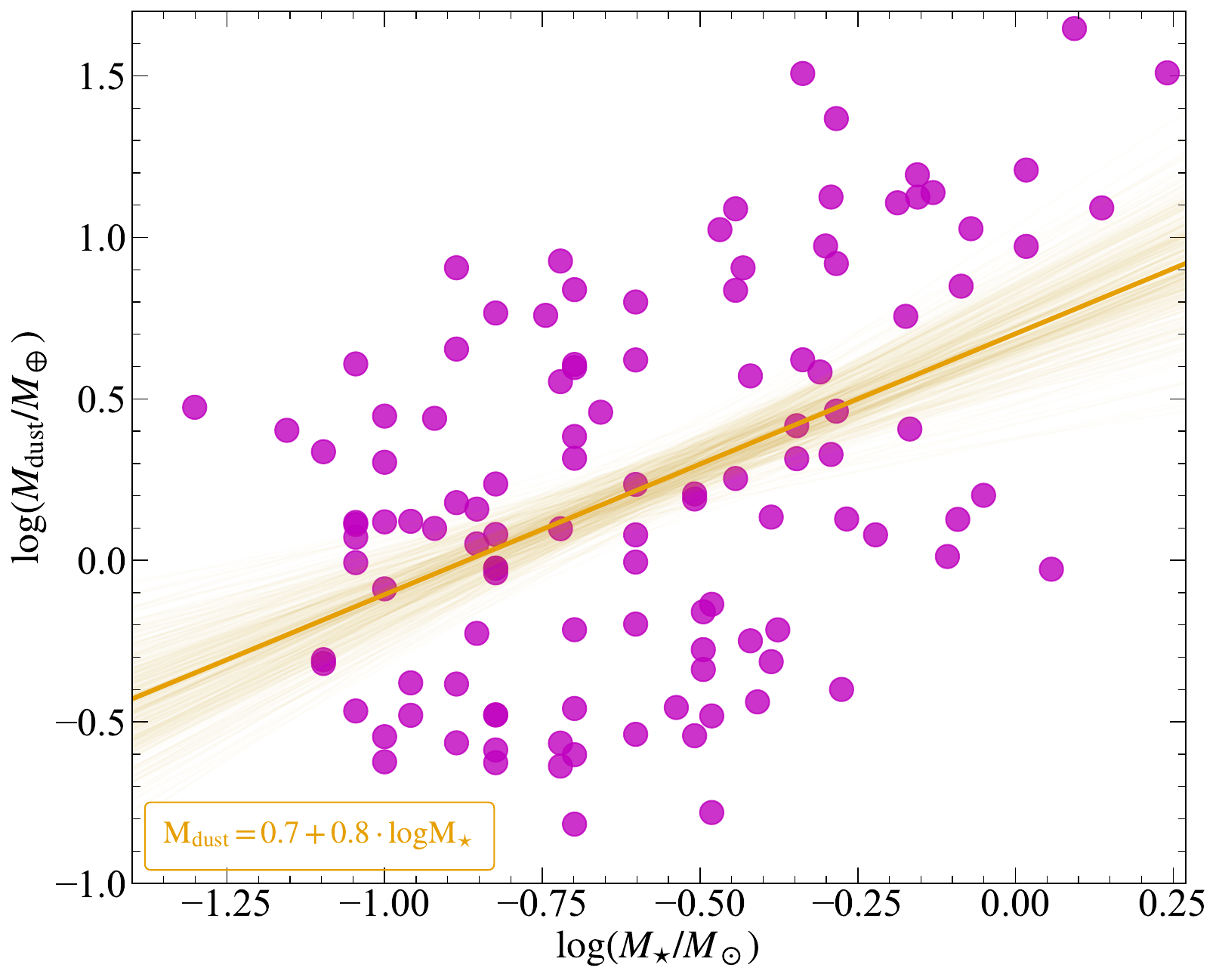}
    \caption{Relation of dust mass to stellar mass. The scatter plot and best fit power law follow the same convention as those in Section \ref{sec: results}. See Table \ref{tab: linmix_params} for the best fit parameters determined by \textit{linmix}.}
    \label{fig: mdust_mstar_linmix}
\end{figure*}

\begin{figure*}
\centering
\includegraphics[width=0.49\textwidth]{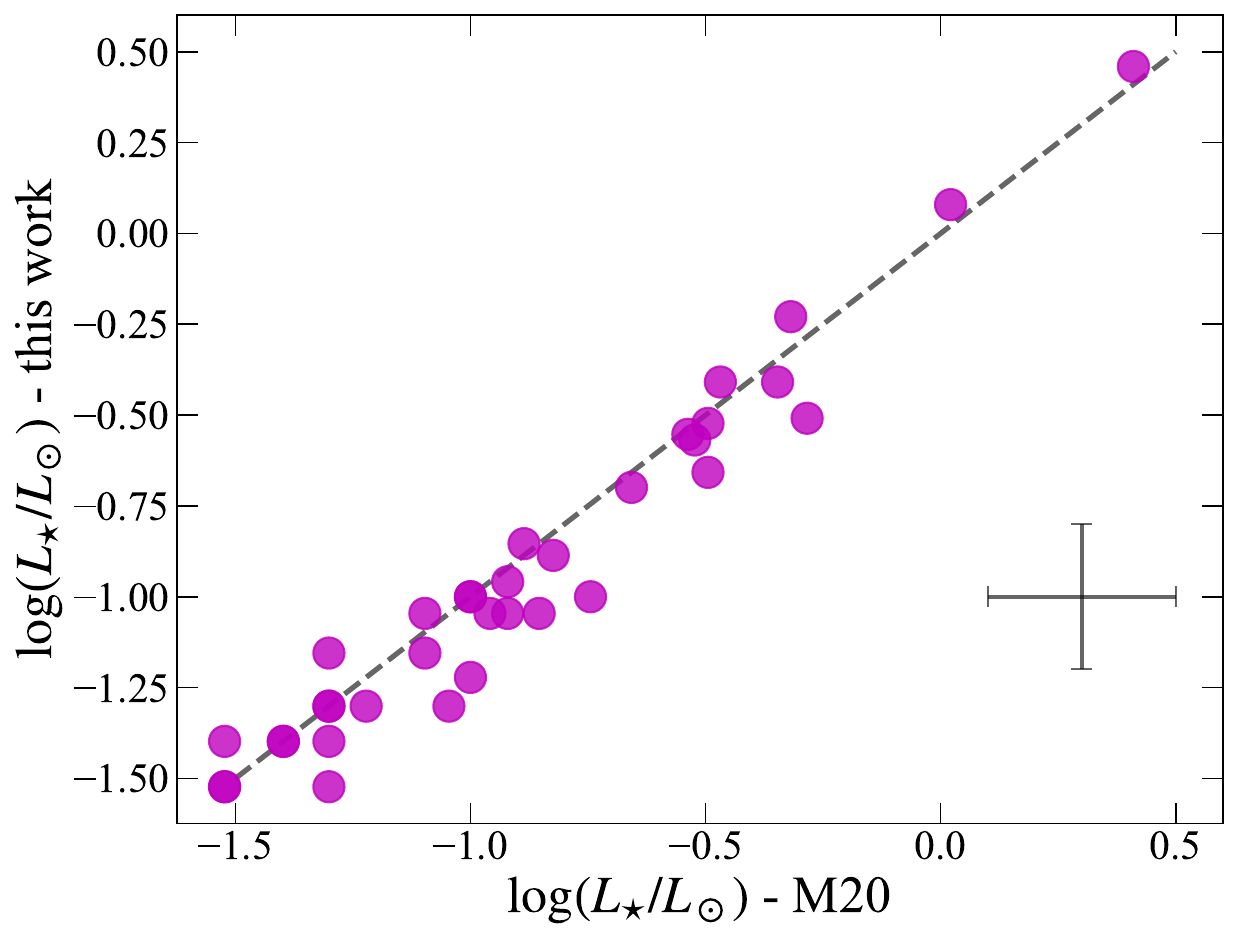}
\hfill
\includegraphics[width=0.49\textwidth]{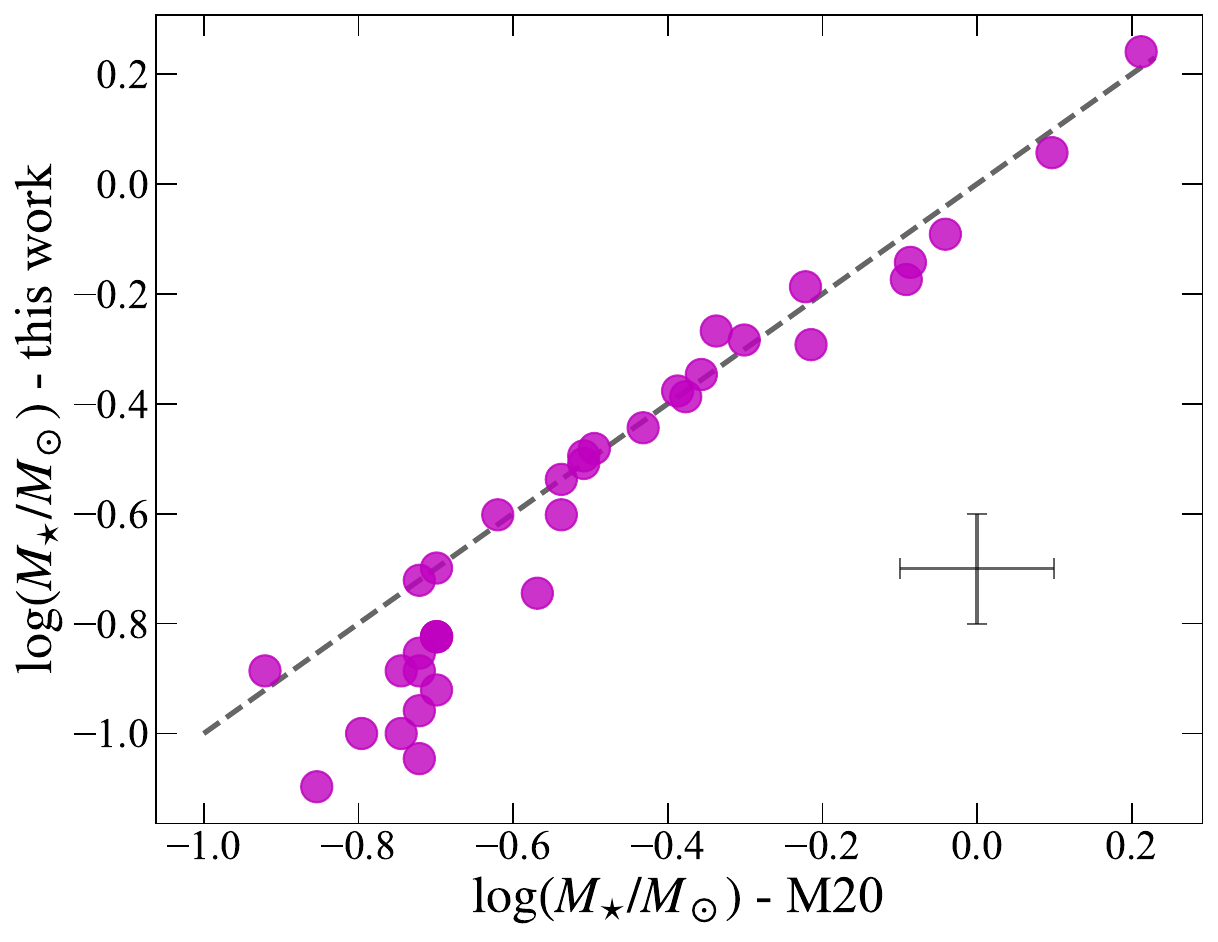}
\vspace{0.05 cm}
\includegraphics[width=0.49\textwidth]{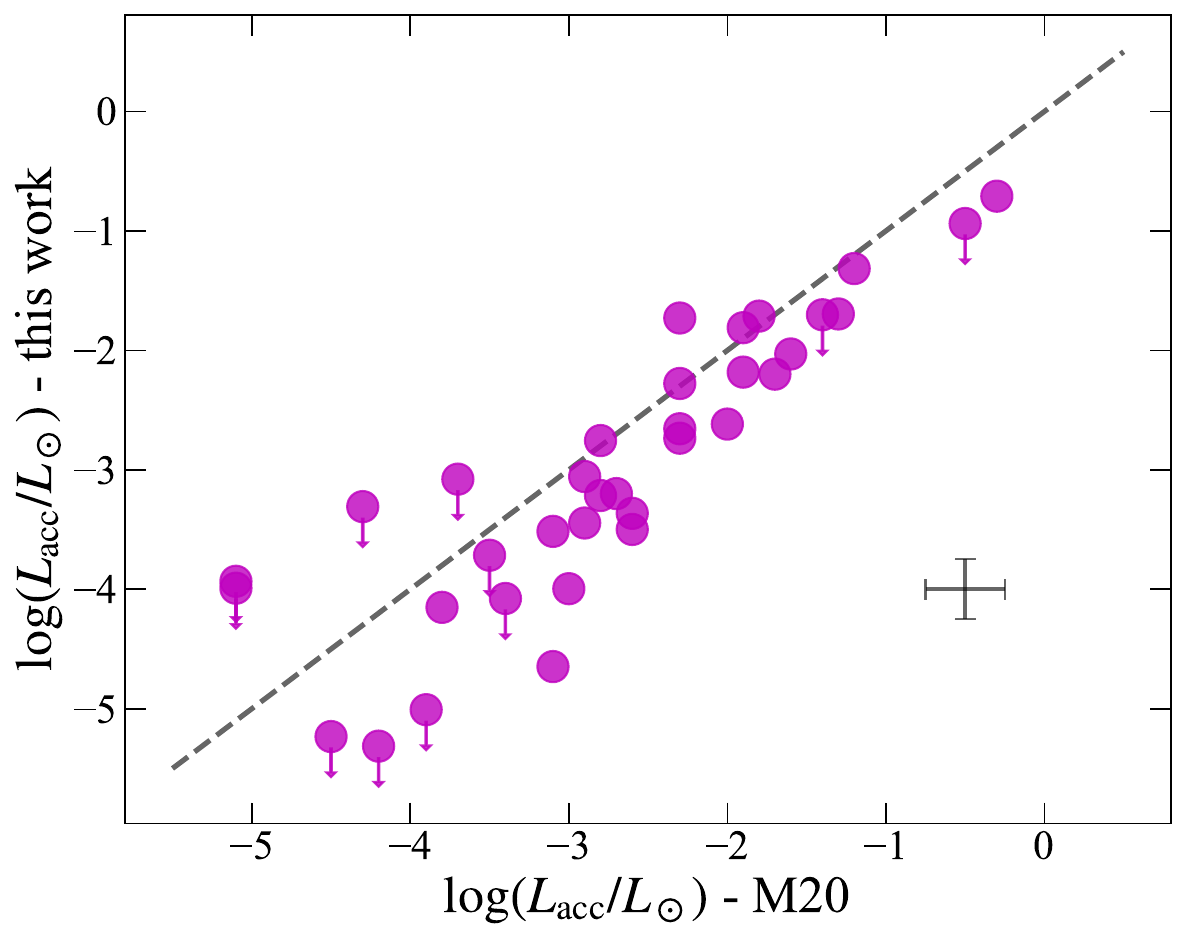}
\hfill
\includegraphics[width=0.49\textwidth]{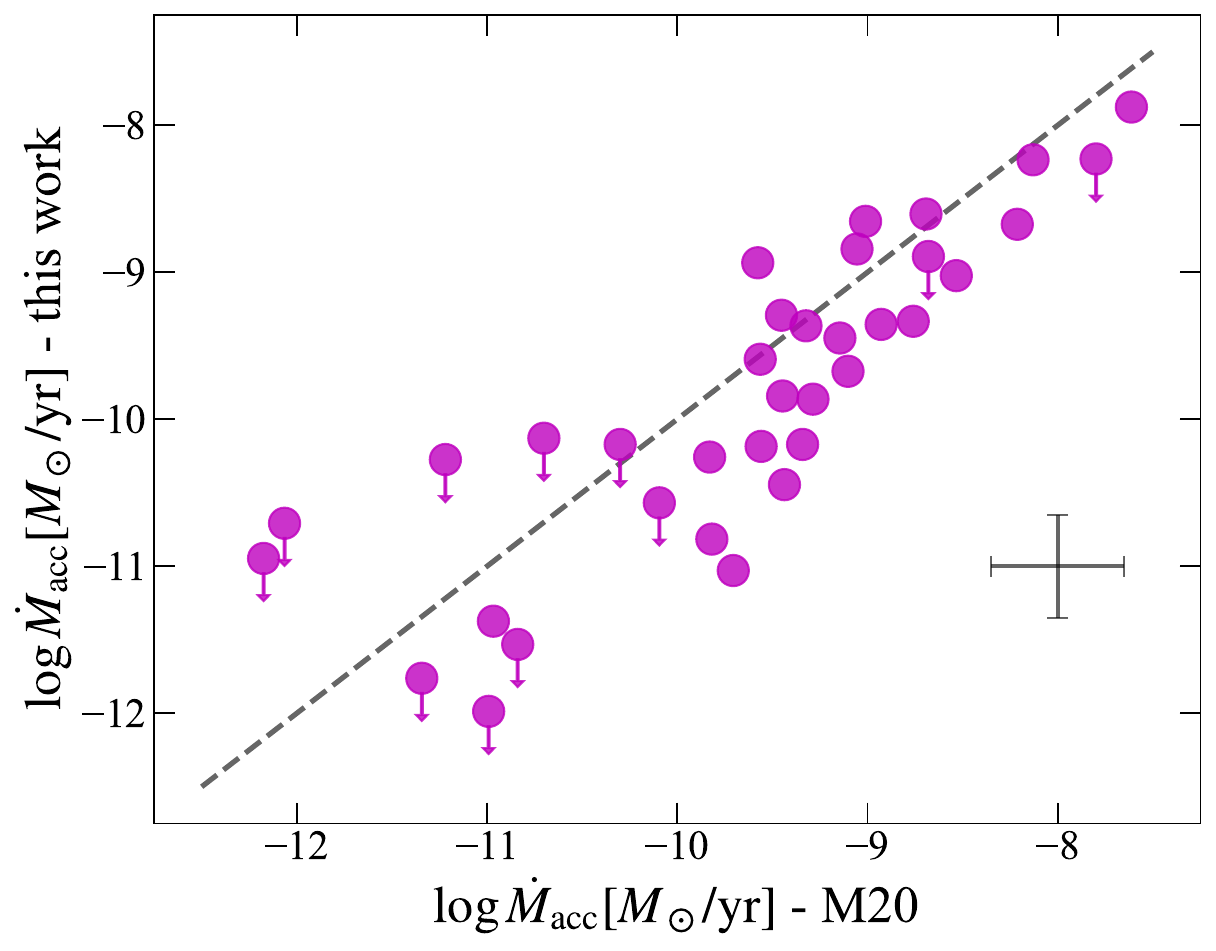}
\vspace{0.05 cm}
\caption{Comparison of \lstar, \mstar, \lacc, and \macc values derived in this work and those derived in M20 for shared objects. Dashed grey lines represent a one-to-one relation. Grey error bars represent the typical uncertainties on the derived values. }
\label{fig: m20_comparison}
\end{figure*}

In Figure \ref{fig: m20_comparison} the values of \lstar, \mstar, \lacc, and \macc ~are compared for shared targets in this work and that of M20. Values are in general in good agreement within uncertainties. Differences in \lacc, \macc are seen particularly in the cases of upper-limit accretors, this can be attributed to the novel approach to determining upper-limit values done in this work. A greater dispersion is also seen when comparing the stellar properties (\lstar, \mstar), particularly at the lowest masses. Such differences stem from the updates to the fitting procedure used in this work (FRAPPE) compared to the previous routine used in M20 (see \citet{Claes2024} for details). Particularly, the use of a different SpT-\teff ~relation can account for the deviations in the stellar mass and luminosity between the two datasets.\\

\FloatBarrier

\onecolumn

\section{Table of results}

\begin{table*}[h]

\caption{Stellar, disc and accretion properties for all objects fit with FRAPPE.}
\centering
\begin{tabular}{@{} c c c c c c c c c c c c c @{}} 
\hline
\hline
2MASS ID             & SpT & Dist  & $T_{eff}$ & $A_V$    & \lstar & \mstar & Age & log\lacc & \macc &\mdust    & UL      & CA \\
       ~             & ~ & [pc]     &  [K]       & [mag]    &   [\lsun]   &    [\msun] & [Myr]&  [\lsun] & [\msun yr$^{-1}$] &    [\msun]& (y/n) & (y/n) \\  
\hline
J16025431-1805300   & M5.0 & 149.3 & 2980 & 0.7 & 0.05  & 0.10  & 2.13   & -4.638  & \num{7.68E-12} & \num{6.04E-06}    & y      & y            \\
J16024152-2138245   & M4.5 & 139.4 & 3085 & 0.4 & 0.03  & 0.13  & 5.40   & -3.445  & \num{6.54E-11} & \num{2.42E-05}    & n      & n            \\
J16140792-1938292   & G9.5 & 159.5 & 4947 & 0.9 & 1.38  & 1.37  & 7.98   & -1.486  & \num{1.52E-09} & \num{3.71E-05}    & y      & n            \\

\multicolumn{13}{c}{\ldots} \\
J16142029-1906481   & K7.5 & 138.8 & 3960 & 2.5 & 0.19  & 0.74  & 15.36  & -0.669  & \num{1.05E-08} & \num{4.13E-05}    &n       &n            \\
\hline
\end{tabular}
\tablefoot{Columns 12 \& 13 refer to if an object is determined to have an upper-limit accretion measurement (UL) or that its \lacc/\lstar\ ratio is below the expected limit for chromospheric activity (CA) }

\par
\raggedright
\vspace{2mm}
(This table is available in its entirety in machine-readable form at the CDS) 
\label{tab: master_props}

\end{table*}

\section{Log of the observations}\label{sec: logobs}

\begin{table*}[h]

\caption{Night log of observations. }
\centering
\begin{tabular}{@{} c c c c c c c c c @{}}
\hline
\hline
2MASS ID          & Date of Observation          & Exp. Time & \multicolumn{3}{c}{Slit Width ['']} & Airmass & I.Q. & SNR \\ \cline{4-6}
                  &   {[}UT{]}                   &  [Nexp x (s)]         & UVB    & VIS    & NIR    &  ["]   &  ["]   & [@ 335 nm] \\
\hline
 J15354856-2958551\_E    &    2018-05-19T23:39:04.373    &    4$\times$300    &    1.0    &    0.4    &    0.4    &    0.42    &    2.22    &    89.8    \\    
   J15354856-2958551\_W    &    2018-05-19T23:39:04.373    &    4$\times$300    &    1.0    &    0.4    &    0.4    &    0.42    &    2.22    &    57.8    \\    
   J15442550-2126408    &    2024-07-25T00:56:53.279    &    4$\times$600    &    1.0    &    0.4    &    0.4    &    0.94    &    1.02    &    166.0    \\    

\multicolumn{9}{c}{\ldots} \\
   J16395577-2347355    &    2024-05-20T09:21:49.414    &    4$\times$360    &    1.0    &    0.4    &    0.4    &    1.14    &    1.65    &    69.2    \\    
\hline
\end{tabular}
\par
\raggedright
\vspace{2mm}
(This table is available in its entirety in machine-readable form at the CDS) 
\label{tab: logobs}

\end{table*}

\end{appendix}

\end{document}